\documentclass[11pt]{article}
\usepackage[T1]{fontenc}
\usepackage{graphicx}    
\usepackage{caption}     

\usepackage{lmodern}
\usepackage{bbm}
\usepackage{amssymb}
\usepackage{amsthm}
\usepackage{amsfonts}
\usepackage{amsmath}
\usepackage{setspace}
\usepackage[utf8]{inputenc} 
\usepackage{amsthm}
\usepackage[numbers]{natbib}
\usepackage[T1]{fontenc}
\usepackage{lmodern,microtype}
\usepackage{graphicx}
\usepackage{titlesec,titling} 
\usepackage[nohead]{geometry} 
\usepackage{setspace} 
\usepackage{amsmath,amsfonts,amssymb,amsthm}
\usepackage{mathrsfs,ushort} 
\usepackage{graphicx,psfrag,epsf}
\usepackage{natbib}
\usepackage{bm}
\usepackage{dsfont}
\usepackage{url} 
\usepackage{hyperref}
\usepackage{bm}
\usepackage{mathtools}
\usepackage{dsfont}
\usepackage{comment}
\usepackage{breqn}
\usepackage{colonequals}
\usepackage{mathtools}
\usepackage{cases}
\usepackage{threeparttable}
\usepackage{mathtools}
\usepackage{graphicx}
\usepackage{empheq}
\usepackage{lmodern}
\usepackage{amsmath}

\usepackage{colonequals}
\RequirePackage{natbib}
\usepackage{titling}
\usepackage{xcolor}
\usepackage{enumitem}
\usepackage{threeparttable, tablefootnote}

\newcommand\Item[1][]{%
	\ifx\relax#1\relax  \item \else \item[#1] \fi
	\abovedisplayskip=0pt\abovedisplayshortskip=0pt~\vspace*{-\baselineskip}}

\DeclareSymbolFont{symbolsC}{U}{pxsyc}{m}{n}
\DeclareMathSymbol{\coloneqq}{\mathrel}{symbolsC}{"42}

\titleformat{\section}[block]{\centering\large\bfseries}{\thesection.}{0.5em}{}
\titleformat{\subsection}[block]{\flushleft\bfseries}{\thesubsection.}{0.5em}{}
\titleformat{\subsubsection}[runin]{\normalsize\itshape}{\bfseries\thesubsubsection.}{0.5em}{}[.--\:]

\usepackage[format=plain,justification=justified]{caption}

\makeatletter
\renewcommand\@biblabel[1]{}
\makeatother

\makeatletter
\@ifundefined{subsubsection}{}{%
	\renewcommand\subsubsection{%
		\@startsection{subsubsection}{3}{\z@}%
		{1.5ex \@plus 1ex \@minus .2ex}
		{0.8ex \@plus .2ex}
		{\normalfont\normalsize\bfseries}
	}%
}
\makeatother

\begin{document}

	\def\spacingset#1{\renewcommand{\baselinestretch}%
		{#1}\small\normalsize}

	

\pretitle{\flushleft\bfseries}
\posttitle{}
\title{\bf Detecting Sparse Cointegration\thanks{Jesus Gonzalo gratefully acknowledges financial support from the {\it Spanish Ministerio de Ciencia e Innovacion}: PID2019-104960GB-IOO; TED2021-129784B-IOO;
CEX2021-001181-M and {\it Spanish Ministerio de Ciencia, Innovacion y Universidades}: PID2023-147593NB-IOO,  financed by MICIU/AEI /10.13039/501100011033.  Jean-Yves Pitarakis gratefully acknowledges financial support from the UK {\it Economic and Social Research Council} through grant ES/W000989/1. The authors are grateful to two anonymous referees for their constructive comments, which significantly improved the paper. }}

\preauthor{\flushleft}
\postauthor{}
\author{%
JES\'{U}S GONZALO\textdagger\ and JEAN-YVES PITARAKIS\textdaggerdbl \\
\bigskip
\textdagger \textit{Department of Economics, Universidad Carlos III de Madrid, C/Madrid 126, 28903 Getafe (Madrid), Spain\\
	(e-mail: jesus.gonzalo@uc3m.es)}\\
\vspace{0.4cm}
\textdaggerdbl \textit{Department of Economics, University of Southampton, UK\\
	(e-mail: j.pitarakis@soton.ac.uk)}
}

	\maketitle

\bigskip
\begin{abstract}
We propose a two-step procedure for detecting sparse cointegration in high-dimensional single-equation models. First, we employ the adaptive lasso to identify the subset of integrated covariates driving the long-run equilibrium relationship with a target series. Second, we adopt an information-theoretic criterion to distinguish between stationarity and nonstationarity in the resulting residuals, avoiding reliance on complex asymptotic distributions. A key theoretical contribution is demonstrating that this residual-based decision rule remains consistent regardless of the internal cointegration structure among the right hand side predictors themselves. Monte Carlo experiments confirm the procedure’s robust finite-sample performance under endogeneity, serial correlation, and rank deficiency in the regressor matrix.
\end{abstract}
	
	\vspace{0.8cm}

	{\it Keywords:}  Cointegration, High Dimensional Data, Adaptive lasso, Unit Roots. 
	
	\medskip
	\noindent
	JEL: C32, C52.
	\vfill
	
	\newpage
	
	\spacingset{1.5}
	
\section{Introduction}

Understanding long-run equilibrium relationships among economic and financial variables is a central part of economic modelling. These relationships often result in time series moving together due to common trends, a phenomenon known as cointegration. Cointegration analysis, which identifies if nonstationary variables share common stochastic trends, has typically been performed in low-dimensional settings. Classical methods for detecting cointegration like the Engle-Granger two-step procedure (Engle and Granger (1987)) or Johansen's maximum likelihood approach (Johansen (1991)) are well-suited for environments with few variables. However, with the advent of high-dimensional data where the number of series of interest to an investigation may be large, traditional methods become inadequate for identifying cointegrating relationships or testing for their presence. In high-dimensional settings, standard estimation techniques like least squares also become unstable, even when computationally feasible. Moreover, the lack of oracle knowledge makes it challenging to identify a small subset of cointegration-inducing variables from a large pool of candidates.

The goal of this paper is to propose a simple method to test for the presence of cointegration among a large pool of I(1) candidate series, particularly in cases where cointegration, if present, is assumed to be sparse and one wishes to identify potential cointegration-inducing covariates. 
The environment is that of a single equation cointegration setting where a target series of interest potentially cointegrates with a small number of series from a large pool of candidates.  The key insight of our approach is that we can detect  cointegration by examining the properties of residuals from a sparse regularized regression.
Consider the problem of determining whether the stock price of a specific constituent of a stock market index, such as the FTSE100 or S\&P500, cointegrates with other stocks in the index. This question holds important implications for portfolio diversification, as cointegrated stocks exhibit stable long-run relationships that mitigate risks. Similarly, it may be of interest to determine whether stocks within similar industries share a common stochastic trend. Such analyses require cointegration methods capable of handling a large number of covariate candidates to detect meaningful relationships. The sparse nature of cointegration that we operate under reflects the idea that most variables in high-dimensional economic and financial data will be irrelevant to any given cointegrating relationship. For example, while a stock's price may cointegrate with a handful of other stocks in the market (e.g., stocks in related industries), it is unlikely to share such a relationship with all 99 other constituents in an index like the FTSE100. 
Identifying a relevant subset of predictors is therefore valuable for interpretation and subsequent analysis. 

A primary challenge in this high-dimensional sparse context is
the selection of a manageable set of candidate covariates that may form a cointegrating relationship.
To address this, our first step employs the adaptive lasso (least absolute shrinkage and selection operator). This method is chosen for its ability to produce parsimonious models while delivering consistent estimates of cointegrating relationships. As we show theoretically and empirically, the adaptive lasso can achieve model selection consistency under favorable conditions or the weaker sure-screening property, and yields super-consistent estimators for non-zero slope coefficients, though these ideal properties may not hold universally when regressors themselves are cointegrated. 

The second critical challenge lies in robustly determining whether the residuals from the relationship estimated in the first step are stationary (I(0)), indicating cointegration, or nonstationary (I(1)), implying a spurious regression.
Although one may be inclined to invoke existing techniques such as ADF type unit root tests (Engle and Granger (1987), Engle and Yoo (1987), Phillips and Ouliaris (1990)) or KPSS type stationarity tests (Kwiatowski et al. (1992), Shin (1994)), both of these face limitations that tend to amplify in high dimensional contexts (e.g., non-standard limiting distributions that depend on the number of fitted covariates requiring model specific tabulations). In addition and as we show in the sequel, under no-cointegration, the adaptive lasso may spuriously select a non-sparse set of predictors (possibly all), making traditional testing based approaches conceptually difficult to design and implement. In this paper we depart from these testing based methods and propose an information-theoretic approach that avoids reliance on asymptotic distribution-based inferences. Crucially, we demonstrate that this residual-based inference remains valid even when variable selection is imperfect as it will typically be the case when the right-hand side covariates are cointegrated themselves. 
An important additional advantage of such an approach is its robustness to phenomena such as endogeneity and serial correlation while also being immune to the the number of fitted covariates. The viewing of inferences about stationarity and non-stationarity as a model selection problem has been explored in the context of unit-root detection in Phillips and Ploberger (1996), Phillips (2008) among others and in the context of vector error correction models in Gonzalo and Pitarakis (1998). 

This work adds to the growing research on high-dimensional estimation and inference in nonstationary settings. Incorporating modern high-dimensional statistical methods into time series analysis is particularly important for understanding economic data. However, the combination of high dimensionality and nonstationarity presents unique challenges. High dimensionality can lead to misleading results in nonstationary environments (e.g. spurious cointegration), as highlighted in Gonzalo and Pitarakis (1999, 2021) and Onatski and Wang (2018). Recent research has developed theoretical tools to address these challenges. Kock (2016) studied the properties of the adaptive lasso in autoregressive models with unit roots, while Koo et al. (2020) and Lee, Shi and Gao (2022) established precise limiting distributions for lasso-based estimators in high-dimensional predictive regressions with unit-root covariates. More recently, Bykhovskaya and Gorin (2022, 2024) and Onatski and Wang (2018, 2021) have developed high-dimensional extensions of Johansen-type maximum eigenvalue and trace tests for no cointegration in large VAR systems, focusing on system-wide inference for the cointegrating rank. Closer to our single-equation perspective, Smeekes and Wijler (2021) proposed an automated sparse cointegration modelling approach based on penalised error-correction models, and Mei and Shi (2024) analysed lasso's properties in high-dimensional predictive regressions with I(1) predictors and a focus on forecasting. 
	
Building on this literature, our paper develops a simple, practical method for detecting sparse cointegrating relationships in a single-equation regression with a potentially large pool of $p$ I(1) covariates, combining adaptive-lasso–based selection of the active set with an information-criterion–based decision on whether the resulting disturbance is I(0) or I(1). Our theoretical analysis characterizes both the ideal behavior of the adaptive lasso under identifiable conditions and its more robust properties for residual-based inference. Our results are developed in a fixed $p$ asymptotic framework, in line with the oracle property literature on adaptive lasso, while our Monte-Carlo experiments explore finite sample behaviour for moderate values of $p$ across various  designs. 

The remainder of the paper is structured as follows. Section 2 introduces the theoretical framework. Section 3 presents the two-step approach to detecting cointegration and derives its theoretical properties. Section 4 discusses practical implementation issues, including penalty selection and finite-sample adjustments. Section 5 reports Monte Carlo results, and Section 6 concludes. A supplementary appendix accompanying this paper contains a selection of additional simulation based outcomes. 

\section{Theoretical Framework}

We consider a single-equation cointegration setting where the target variable \( y_t \) may cointegrate with a subset of predictors drawn from a large pool of \( p \) I(1) series. The operating model is given by
\begin{align}
	y_t & = \beta_0 + \bm{x}_t' \bm{\beta} + z_t, \quad t = 1, \ldots, n,
	\label{eq:eq1}
\end{align}
where \( y_t \in \mathbb{R} \) is the target variable, \( \bm{x}_t = (x_{1t}, x_{2t}, \ldots, x_{pt})' \in \mathbb{R}^p \) is a \( p \)-dimensional vector of covariates, \( \bm{\beta} = (\beta_1, \beta_2, \ldots, \beta_p)' \in \mathbb{R}^p \) is the vector of unknown slope coefficients, \( \beta_0 \in \mathbb{R} \) is the intercept term, and \( z_t \in \mathbb{R} \) represents the deviation from the equilibrium relationship.

The covariates \( \bm{x}_t \) are modeled as I(1) processes, possibly correlated across dimensions and with serially correlated disturbances
\begin{align}
	x_{jt} & = x_{j,t-1} + v_{jt}, \quad j = 1, \ldots, p, \quad t = 1, \ldots, n.
	\label{eq:eq2}
\end{align}
The deviation term \( z_t \) is modelled as
\begin{align}
	z_t & = \rho z_{t-1}+u_{t}, \quad t = 1, \ldots, n.
	\label{eq:eq3}
\end{align}
Letting $\bm \eta_{t}=(u_{t},v_{1t},\ldots,v_{pt})'$, we model these $p+1$ disturbance series as
\begin{align}
\bm \eta_{t} & = \bm C(L) \ \bm e_{t}
	\label{eq:eq4}
\end{align}
where $\bm C(L)=\sum_{i=0}^{\infty}C_{i}e_{t-i}$ for $\sum_{i=0}^{\infty}i |\bm C_{i}|<\infty$, $\bm C_{0}=I$ and ${\bm e}_{t} \sim i.i.d.(0,\bm \Sigma_{e})$ with $E\|\bm e\|^{2+\delta}<\infty$ for some positive $\delta$. These assumptions are standard in this literature and essentially ensure that an FCLT holds for the ${\bm \eta}_{t}$ sequence (see Phillips and Durlauf (1986)). Under cointegration, the existence of a long-run equilibrium relationship implies that \( z_t \) is stationary with $|\rho|<1$, even though both \( y_t \) and the components of \( \bm{x}_t \) are individually I(1) processes. If $\rho=1$ instead, (\ref{eq:eq1}) is viewed as a spurious regression.  
By ``spurious regression'' under $\rho=1$ we mean that there is no vector $\bm b \in \mathbb{R}^p$ such that $y_t - \bm x_t' \bm b$ is $I(0)$, i.e.\ $y_t$ is not cointegrated with any linear combination of the regressors.

To formalise the notion of sparse cointegration, let
\begin{align}
S & = \{ j \colon \beta_j \neq 0 \}, \quad j = 1, \ldots, p
	\label{eq:eq5}
\end{align}
denote the set of active covariates inducing cointegration, with \( |S| = s \ll p \). The remaining covariates with \( \beta_j = 0 \) are irrelevant for the cointegrating relationship. Note that throughout this paper we denote the true parameter vector by $\bm\beta^0$ and write $\beta_j^0$ for its $j$th component. When there is no ambiguity, we may drop the superscript and write simply $\bm\beta$.
The notion of sparse cointegration posits that the target variable \( y_t \) is cointegrated with only a small number of covariates among the large pool of \( p \) candidates. Formally, this is expressed by assuming that the cardinality of the active set \( S \) is much smaller than \( p \), i.e., \( s \ll p \). This assumption reflects the realistic scenario where most covariates are irrelevant for the equilibrium relationship, as often encountered in applications with high-dimensional economic and financial data. Sparse cointegration implies that \( \bm{\beta} \) is a sparse vector, where
\begin{align}
\beta_j & \neq 0 \quad \text{if and only if } j \in S, \quad \beta_j = 0 \quad \text{for } j \notin S.
	\label{eq:eq6}
\end{align}

In line with the above description of the notion of sparse cointegration, we let the $s \times 1$ vector ${\bm \beta}_{S}$ collect the parameters whose covariates induce cointegration, when the latter is present. Similarly we let ${\bm x}_{{S},t}$ denote the $s-vector$ of active covariates associated with ${\bm \beta}_{S}$. The parameters associated with the variables that do not actively enter  (\ref{eq:eq1}) are in turn collected in the $(p-s)$ vector
${\bm \beta}_{{S}^{c}}$ while ${\bm x}_{{S}^{c},t}$ collects the $(p-s)$ inactive series. 

Our goal is to assess whether (\ref{eq:eq1}) is truly a cointegrating relationship by analyzing the stationarity of a sample counterpart of $z_{t}$ when $p$ is large and only a small unknown number of these series induce cointegration, if the latter is truly present. If sparse cointegration is present for instance
we expect the adaptive lasso based residuals to behave like an I(0) process which our approach will be designed to detect. Similarly, if the regression is spurious with $\rho=1$ we expect these residuals to behave like an I(1) process. 

Throughout this paper we adopt a fixed-$p$ asymptotic framework with $n \to \infty$. This choice is in line with the oracle-property literature on adaptive lasso (e.g., Zou (2006), Kock (2016)) and allows us to obtain transparent conditions under which our two-stage procedure behaves as intended. 
We do not claim that fixed-$p$ asymptotics are the only or universally preferable framework in high-dimensional settings. Recent contributions also focused on 
establishing increasing-dimension asymptotics for lasso-type estimators in nonstationary environments. In Smeekes and Wijler (2021) and Mei and Shi (2024) for instance, the authors developed complementary high-dimensional asymptotic theories in which $p$ is allowed to grow with $n$, typically under different objectives (e.g., automated ECM specification or predictive regression based inferences).
Our notion of “high dimensionality” here refers instead to the large {\it model space} generated by $p$ potential regressors. Even for moderate values of $p$, the number of possible models grows as $2^{p}$, rendering exhaustive or stepwise classical selection methods impractical. Moreover, it is well known that ordinary least squares becomes unstable when $p/n$ is large, even if $p < n$, and that estimating a large number of slope coefficients by least squares typically leads to residuals that accumulate substantial estimation error and are therefore noisy objects for deciding about cointegration. In this sense, our fixed-$p$ asymptotic analysis should be viewed as a tractable baseline that captures the sparse-cointegration structure and is complemented by Monte Carlo experiments that explore finite-sample performance in moderate-$p$ settings that are empirically relevant.

\section{Estimation and Testing}

We initially motivate and describe the use of the adaptive lasso as a method to perform model-selection and estimation in a single shot within the cointegrating regression model in (\ref{eq:eq1}). The adaptive lasso estimator of the model parameters is given by
\begin{align}
(\hat{\beta}_0^{AL},\hat{\bm \beta}^{AL}) & = \arg\min_{\beta_0,\bm \beta} \Biggl\{ \sum_{t=1}^n (y_t - \beta_0 - \bm x_t' \bm \beta)^2 + \lambda_n \sum_{j=1}^p w_j|\beta_j| \Biggr\}.
\label{eq:eq7}
\end{align}
where $\lambda_{n}$ is a regularization parameter (penalty term) and the $w_{j}$'s are the adaptive weights associated with the $\beta_{j}'s$. Specifically, given an initial estimator $\widetilde{\beta}_{j}$ (e.g., OLS) these weights are set as $w_{j}=1/|\widetilde{\beta}_{j}|^{\gamma}$ with $\gamma>0$. The intercept $\beta_0$ is not penalised in \eqref{eq:eq7}. 

The adaptive lasso is particularly well-suited to the cointegration setting considered in this paper, as it addresses several limitations of the standard lasso. In high-dimensional regression problems, the standard lasso suffers from shortcomings that are particularly detrimental to our goal of consistently estimating the residuals of a sparse cointegrating regression. Unless strong assumptions are imposed (e.g., irrepresentable condition), it does not satisfy the oracle property, which guarantees asymptotic consistency and correct model selection. The adaptive lasso aims to resolve these issues by introducing data-dependent weights that penalize small coefficients more heavily, with the goal of allowing for consistent estimation of large coefficients while effectively shrinking irrelevant predictors to zero.
For instance, if $\widetilde{\beta}_{j}$ is near zero, the associated $w_{j}$ will be large, which can help penalize parameters associated with variables that are unlikely to be active. 

Unlike the standard lasso, which imposes the same penalty on all coefficients regardless of their magnitude, the adaptive lasso adjusts the penalty weights based on an initial estimate of the coefficients. By applying smaller penalties to coefficients with larger initial estimates, the adaptive lasso allows these coefficients to converge more accurately to their true values, while continuing to shrink the irrelevant coefficients to zero. This feature makes the adaptive lasso particularly suitable for constructing residuals $\hat{z}_{t}$ that mimic the true $z_{t}'s$. 

Before proceeding further it is also important to discuss the explicit inclusion of an intercept term and the use of raw (unstandardised) I(1) series in \eqref{eq:eq7}. In typical lasso applications with stationary (I(0)) data, it is common practice to standardize predictors to have zero mean and unit variance. This ensures that the $\ell_{1}$ penalty is applied on a comparable scale across coefficients. For nonstationary I(1) series, however, standard ``standardisation'' (subtracting a sample mean and dividing by a sample standard deviation) is problematic. Their sample mean and variance are not well-defined in the usual sense, and such a transformation would also complicate the interpretation of cointegrating coefficients, which are defined in terms of the original scales of the variables. Moreover, in a cointegration setting like ours, the intercept $\beta_{0}$ is part of the long-run equilibrium relationship and is not merely a nuisance parameter. Our model in \eqref{eq:eq1} includes $\beta_{0}$ and estimates it as part of the adaptive lasso procedure outlined in \eqref{eq:eq7}. Crucially, the penalty term $\lambda_{n}\sum_{j=1}^{p}w_{j}|\beta_{j}|$ applies only to the slope coefficients $\beta_{j}$ and not to the intercept. The Karush–Kuhn–Tucker (KKT) conditions ensure that the residuals sum to zero, analogous to ordinary least squares, and by the Frisch–Waugh–Lovell theorem (see Yamada (2017)) this is equivalent to running the adaptive lasso on demeaned variables while keeping the intercept unpenalised. 

While the standard lasso is sensitive to the scale of predictors, the adaptive lasso mitigates this issue via its data-based weights, which automatically adjust penalties based on the apparent importance of each variable.
The adaptive weights $w_{j}$ adjust the penalty for each coefficient. If an initial estimate $\widetilde{\beta}_{j}$ is large (which may reflect a strong underlying effect and/or the scale of $x_{jt}$), its corresponding weight will be small, leading to less shrinkage for that coefficient; conversely, a small $\widetilde{\beta}_{j}$ leads to a large weight and thus stronger shrinkage. These adaptive data-dependent weights therefore absorb much of the variation in the magnitudes and scales of the I(1) predictors and mitigate the need for prior standardization purely for the purpose of equitable penalization. In what follows, the theoretical properties of the adaptive lasso, including model selection consistency and $n$-consistency of the estimated active slope parameters, are derived in the context of such I(1) regressors without prior standardization.

Locating the optima of a function such as (\eqref{eq:eq7}) is a convex optimisation problem which guarantees that any local minimum is also a global minimum. The structure of the program however is challenging due to the inclusion of the $\ell_{1}$ norm penalty in the objective function. Note for instance that although the $\ell_{1}$ norm is convex, it is non-differentiable at points where $\beta_{j}=0$. This creates difficulties for standard optimisation techniques such as gradient descent which relies on smoothness. Instead, coordinate descent type of algorithms which avoid the need to compute a full gradient at non-differentiable points are typically considered. 

Given $\hat{\bm \beta}^{AL}$ estimated from \eqref{eq:eq7}, we define
\begin{align}
	\hat{S} & = \{j \colon \hat{\beta}_{j}^{AL}\neq 0\}
	\label{eq:eq8}
\end{align}
as the active set of predictors selected by the adaptive lasso, and in the sequel refer to the covariates associated with this estimated active set as $\bm{x}_{\hat{S},t}$. We next use this estimated active set $\hat{S}$ to form the residuals of interest. These can be obtained using the adaptive lasso estimates directly
\begin{align}
	\hat{z}_{t} & = y_{t}-\hat{\beta}_{0}^{AL}-\bm{x}_{t}'\hat{\bm \beta}^{AL},
	\label{eq:eq9}
\end{align}
noting that the vast majority of the components of $\hat{\bm \beta}^{AL}$ are set to zero by the effect of the $\ell_{1}$ penalization. Under sparse cointegration with $|\rho|<1$, these residuals behave as I(0) processes, while under $\rho=1$ they behave as I(1) processes. Thus for our second-stage procedure, they serve as valid proxies for $z_t$ in terms of their order of integration properties 

Given the residual sequence $\hat{z}_{t}$ constructed as above, our next goal is to assess whether or not it contains a unit root in its autoregressive representation. For this purpose consider the following two competing models
\begin{align}
\Delta \hat{z}_{t} & = \mu + \sum_{j=1}^{k} \phi_{j}\Delta \hat{z}_{t-j}+\epsilon_{t} 	\ \ \ \textrm{model} \ \  {\cal M}_{0} \label{eq:eq10} \\
\Delta \hat{z}_{t} & = \mu + \phi \ \hat{z}_{t-1}+\sum_{j=1}^{k} \phi_{j}\Delta \hat{z}_{t-j}+\epsilon_{t} \ \ \ \textrm{model} \ \ {\cal M}_{1}.	\label{eq:eq11}
\end{align}

\noindent
Model ${\cal M}_{0}$ imposes I(1) behaviour by fitting an AR($k$) process in first differences to the residuals. In contrast, model
${\cal M}_{1}$ augments this specification with a lagged level $\hat{z}_{t-1}$ and allows the $\hat{z}_{t}$'s to be stationary when $\phi<0$. Because the $\hat{z}_{t}$'s are centered by construction, the inclusion of an intercept term is not crucial for our asymptotic arguments. The order $k$ of the autoregressions is directly linked to the behaviour of the error process $u_{t}$ driving the equilibrium errors as formulated in \eqref{eq:eq3}. 

We now view the objective of testing for cointegration as a model selection problem between models ${\cal M}_{0}$ and ${\cal M}_{1}$.
Selecting model ${\cal M}_{0}$ indicates absence of cointegration (i.e. \eqref{eq:eq1} is a spurious regression).
In contrast, support for model ${\cal M}_{1}$ implies that the $\hat{z}_{t}$'s behave like an I(0) process so that $y_{t}$ and the set of covariates selected by the adaptive lasso are cointegrated.
Letting $\overline{\sigma}^{2}_{0}$ and $\overline{\sigma}^{2}_{1}$ denote the residual variances from (\ref{eq:eq10}) and (\ref{eq:eq11}), selection between the two models is made using the criteria
\begin{align}
IC_{n,0}(\hat{z}) & = \ln \overline{\sigma}^{2}_{0} + \frac{c_{n}}{n} \ (k+1) \label{eq:eq12} \\
IC_{n,1}(\hat{z}) & = \ln \overline{\sigma}^{2}_{1} + \frac{c_{n}}{n} \ (k+2) \label{eq:eq13}
\end{align}
where $c_{n}$ is a penalty term and $k\in \{0,1,\ldots,k_{max}\}$. The proposed model-selection based approach is based on comparing $IC_{n,0}(\hat{z})$ with $IC_{n,1}(\hat{z})$ and leads to choosing ${\cal M}_{0}$ if $IC_{n,0}(\hat{z})\leq IC_{n,1}(\hat{z})$ and to choosing ${\cal M}_{1}$ if $IC_{n,0}(\hat{z})>IC_{n,1}(\hat{z})$.
Note that this model selection based approach bears strong resemblance with a likelihood ratio type test since
the requirement $IC_{n,0}(\hat{z})>IC_{n,1}(\hat{z})$ for {\it rejecting} ${\cal M}_{0}$ reduces to $n \ln \overline{\sigma}^{2}_{0}/\overline{\sigma}^{2}_{1}>c_{n}$. Here the penalty term $c_{n}$ plays a similar role to the critical values used to form the rejection region of likelihood ratio type statistics, although we emphasise that our analysis proceeds entirely via information criteria rather than the null distribution of a test statistic. It is indeed useful to observe that the above decision based inferences essentially rely on the sign of $\Delta IC_{n}(\hat{z}) \coloneqq IC_{n,0}(\hat{z})-IC_{n,1}(\hat{z})$, $\Delta IC_{n}(\hat{z}) \gtrless 0$, rather than its actual distribution. \\

In summary, our two-stage procedure for detecting sparse cointegration operates as follows:
\begin{itemize}
	\item[] {\bf \sf Stage 1 (Regularized Estimation):} The adaptive lasso is applied to the high-dimensional regression to select a parsimonious set of candidate covariates and estimate the cointegrating vector. The resulting residuals $\hat{z}_{t}$ are constructed.
	\item[] {\bf \sf Stage 2 (Residual-Based Inference):} The stochastic properties of $\hat{z}_{t}$ are assessed using an information-theoretic model selection criterion to discriminate between an I(0) (stationary) and an I(1) (unit root) specification. A finding of I(0) residuals indicates the presence of a cointegrating relationship among the selected variables.
\end{itemize}

\noindent
\paragraph{Remark 1.} In \eqref{eq:eq10} and \eqref{eq:eq11} the order $k$ of the fitted autoregressions is meant to capture the potential presence of serial correlation in the $u_{t}$'s driving the equilibrium errors. In classical unit-root and cointegration testing, such augmentation is fundamental for achieving nuisance-parameter-free asymptotics for test statistics. In our model-selection context, however, the asymptotic ability of the proposed procedure to distinguish between ${\cal M}_{0}$ and ${\cal M}_{1}$ hinges on the relative orders of the residual sums of squares under I(0) and I(1) specifications, and not on the detailed limiting distribution of any test statistic. Heuristically, if $c_{n}\rightarrow \infty$, correct decisions will typically be ensured provided that $n \ln (\overline{\sigma}^{2}_{0}/\overline{\sigma}^{2}_{1})$ is $O_{p}(1)$ under the appropriate model, regardless of whether its limit is characterised by nuisance parameters induced by serial correlation and/or endogeneity. Nevertheless, the inclusion of a sufficiently rich lag structure will favourably influence finite-sample properties.

\subsection{Theoretical Properties}

\subsubsection{\textbf{Asymptotic Properties of the Adaptive Lasso (Stage 1): Consistency and Support Recovery}} 

We first analyze the adaptive lasso's behaviour in ideal conditions 
to gain intuition on its characteristics within our  non-stationary environment. Our analysis requires  distinguishing across alternative structures on the cointegration properties of $\bm{x}_{t}$ itself, beyond the generic framework introduced in Section~2 and which remained silent about the presence or absence of cointegration within the ${\bm x}_{t}'s$ themselves. 

Our propositions below are understood to hold under the stochastic framework described in Section~2 and more specifically \eqref{eq:eq1}-\eqref{eq:eq4}. For the purpose of establishing the selection and estimation properties of the 
adaptive lasso under $|\rho|<1$ and $\rho=1$ (Propositions 1 and 2) we will need to operate under additional conditions on the cointegration properties of the ${\bm x}_{t}'s$ themselves, akin to the single-equation environments considered in the classical Engle-Granger type of analyses or the predictive regression literature.  
Crucially however, our core methodological contribution 
on the actual detection of sparse cointegration via the proposed model selection approach based on \eqref{eq:eq10}-\eqref{eq:eq13} (Proposition 3) in Stage 2 will be seen to remain valid regardless of whether the right hand side components are cointegrated or not (including the possibility of cointegration within the active predictors themselves). Differently put, the environment under which our Proposition 3 holds can be much weaker than the restrictions we impose in Propositions 1 and 2.  \\

\noindent
\emph{Proposition 1}. \emph{Suppose that $y_t$ and $\bm{x}_t$ satisfy the framework of Section~2 with $|\rho|<1$. Let $S = \{j:\beta^0_j\neq 0\}$ denote the active index set,
	and let $\widehat{\bm\beta}^{AL}$ be the adaptive lasso estimator based on the
	full regression of $y_t$ on $(\bm x_t)_{t=1}^n$ with weights
	$w_j = |\widetilde{\beta}^{\,ols}_j|^{-\gamma}$, $\gamma>0$, where $\widetilde{\bm\beta}^{\,ols}$
	denotes the full-sample OLS estimator}.

\begin{enumerate}[label=(\alph*)]
	\item \emph{Assume that the active regressors ${\bm x}_{S,t}$ 
		form a linearly independent I(1) system in the sense that no nonzero linear combination of ${\bm x}_{S,t}$ is I(0), and that no nonzero linear combination involving at least one active regressor and any number of inactive regressors is I(0); the inactive block ${\bm x}_{S^{c},t}$ may be internally cointegrated. Suppose further that the nonzero coefficients satisfy a beta-min condition
	$\min_{j\in S}|\beta^0_j|\ge \underline b>0$ for some $\underline b>0$ and that the penalty sequence $\lambda_{n}$ is such that}
\begin{equation}
	\lambda_n \to \infty,
\qquad
\frac{\lambda_n}{n} \to 0
\qquad\text{as }n\to\infty. \nonumber
\end{equation}

\emph{Then, with $p$ fixed and as $n\to\infty$}

\begin{enumerate}[label=(\roman*)]
	\item \emph{The adaptive lasso estimator is n-consistent on the active set $n\big(\widehat{\bm\beta}^{AL}_S - \bm\beta^0_S\big) = O_p(1)$.}

	\item \emph{The adaptive lasso has the sure-screening property
	$P\big(S \subseteq \widehat S\big) \;\to\; 1$, and
	no truly active regressor is dropped asymptotically.}
\end{enumerate}

\item \emph{Assume in addition that there is no internal cointegration among the regressors as a whole, in the sense that every nonzero linear combination of ${\bm x}_{t}$ is I(1). Suppose that the penalty sequence satisfies, besides the conditions in part (a),}
\begin{equation}
	\frac{\lambda_n}{n^{1-\gamma}} \to \infty \qquad\text{as }n\to\infty. \nonumber
\end{equation}
\emph{Then, with $p$ fixed and as $n\to\infty$}

\begin{enumerate}[label=(\roman*)]
\item \emph{The adaptive lasso estimator achieves full model selection consistency, $P(\widehat S = S) \;\to\; 1$, and the super-consistent rate
$n(\widehat\beta^{AL}_j - \beta^0_j)=O_p(1)$ continues to hold for all $j \in S$, as in \emph{a(i)}}
\end{enumerate}

\end{enumerate}

Proposition~1 collects a set of basic properties of the adaptive lasso in the cointegrating regression with $|\rho|<1$. We note for instance that 
on $S$ the adaptive lasso behaves like a version of the OLS estimator and has the same super-consistency as in the unpenalised cointegrating regression. This result holds 
even if the covariates in ${\bm x}_{S^{c},t}$ are themselves cointegrated, but requires ruling out  any form of cointegration involving the components of ${\bm x}_{S,t}$ as in the classical cointegration literature. This is to ensure that the support estimated by the adaptive lasso contains at least all of the truly active covariates. Part~a(ii) of Proposition 1 formalises a ``sure screening'' property. With probability tending to one, no truly active regressor is dropped from the selected model $\hat S$. If cointegration within the ${\bm x}_{t}'s$ is entirely ruled out however and an additional condition holds on the penalty sequence, then full model selection consistency is achieveable as stated in part \emph{(b)(i)} of the Proposition. 

\paragraph{Remark 2.}
The intuition behind the model selection consistency and its weaker sure-screening counterpart of the adaptive lasso is best understood from the formulation of the KKT conditions associated with the objective function in (\ref{eq:eq7}). Let
\[
S_{n,j} \;\coloneqq\; \frac{1}{n}\sum_{t=1}^n x_{j,t}\,\widehat z_t
\]
denote the normalised score component for regressor $j$, where $\widehat z_t$ are the adaptive lasso residuals. At any adaptive lasso solution the KKT conditions require that for each coordinate $j$
\[
|S_{n,j}|
\;\le\; \frac{\lambda_n}{2 n} w_j
\quad\text{whenever }\widehat\beta^{AL}_j = 0.
\]
Intuitively
a regressor can be excluded from the selected model only if its sample
covariance with the residuals is dominated by the adaptive penalty.  When right hand side cointegration is excluded as in part (b) of Proposition 1, we have for $j \in S$, $S_{n,j}=O_{p}(1)$ with its specific continuous asymptotic distribution containing no atom at zero. At the same time $w_{j}=O_{p}(1)$, and $\lambda_{n}w_{j}/n\rightarrow 0$ so that the KKT condition fails and $P(\hat{\beta}^{AL}_{j}=0)\rightarrow 0$. This is the sure screening property whereby active coefficients cannot be set to zero. For the inactives with $j \in S^{c}$, we also have $S_{n,j}=O_{p}(1)$ but $w_{j}=O_{p}(n^{\gamma})$ and the requirement $\lambda_{n} n^{\gamma-1}\rightarrow \infty$ comes into play. Thus the KKT condition holds and all inactives are dropped, yielding full model selection consistency as stated in part (b) of Proposition 1.  \\

Before proceeding with the stage 2 inferences we also document the adaptive lasso's behaviour in a spurious regression setting (i.e., under $\rho=1$). This is summarized in Proposition 2 below restricted to an environment where the right hand side regressors are assumed not to be cointegrated. \\

\noindent
\emph{Proposition 2}. \emph{Suppose that $y_t$ and $\bm{x}_t$ satisfy the framework of Section~2 with $\rho=1$ in (\ref{eq:eq3}), so that $z_{t}$ is a random walk and (\ref{eq:eq1}) is a spurious regression in the sense that $y_{t}$ is not cointegrated with any linear combination of ${\bm x}_{t}$. Let $S = \{j:\beta^0_j\neq 0\}$ denote the active index set with
	$|S|=s$ fixed, and let $\widehat{\bm\beta}^{AL}$ be the adaptive lasso estimator based on the
	full regression of $y_t$ on $(\bm x_t)_{t=1}^n$ with weights
	$w_j = |\widetilde{\beta}^{\,ols}_j|^{-\gamma}$, $\gamma>0$, where $\widetilde{\bm\beta}^{\,ols}$
	denotes the full-sample OLS estimator. Assume that every non-zero linear combination of the ${\bm x}_{t}'s$ is I(1) i.e. there is no cointegration  within the active block, within the inactive block,  or between the two blocks. Suppose further that the penalty sequence $\lambda_n$ satisfies
\begin{equation}
	\lambda_n \to \infty,
	\qquad
	\frac{\lambda_n}{n} \to 0
	\qquad\text{as }n\to\infty. \nonumber
\end{equation}
\noindent
Then, with $p$ fixed and $n \rightarrow \infty$, we have (i) $\hat{\bm \beta}^{AL}=O_{p}(1)$, and (ii) for each coordinate $j \in \{1,2,\ldots,p\}$, $P(\hat{\bm \beta}^{AL}_{j}=0)\rightarrow 0$ , and hence with $p$ fixed $P(\hat{S}=\{1,2,\ldots,p\})\rightarrow 1$, $\hat{S} \coloneqq \{ j: \hat{\beta}_{j}^{AL}\neq 0 \}$
} \\

Proposition 2 establishes that the adaptive lasso asymptotically selects all regressors in the spurious regression with a full rank I(1) regressor system (no right hand side cointegration). The penalty cannot shrink any coefficient exactly to zero and the adaptive lasso essentially behaves much like OLS in terms of support selection. 

\paragraph{Remark 3.}
Our result in Proposition 2 has ruled out any type of cointegration within the ${\bm x}_{t}'s$ in the right hand side. 
We do not characterize the limiting support of the adaptive lasso in such instances. {\it Internal} cointegration among inactive regressors as considered in Proposition 1(a) for instance would create stationary directions within the design matrix, and we may expect that the 
adaptive lasso will eliminate redundant predictors even under spurious regression ($\rho=1$) to minimize the weighted penalty. 
Consequently, while the selected set $\hat{S}$ is likely to remain large, it need not converge to the full set $\{1,\dots,p\}$.
Our Proposition 3 below establishes that the resulting residuals $\hat{z}_t$ consistently inherit the true order of integration of $z_t$ (behaving as $I(0)$ under cointegration and $I(1)$ under spurious regression) regardless of this internal regressor structure.

\subsubsection{\textbf{ADF regression based model selection and sparse cointegration detection (Stage 2)}} 

We now focus on detecting whether the estimated residuals $\{\hat z_t\}$ are
consistent with an $I(0)$ or an $I(1)$ process under $|\rho|<1$ and $\rho=1$ respectively.
Recall from \eqref{eq:eq12}–\eqref{eq:eq13} that our model–selection rule compares
two ADF–type specifications for $\hat z_t$:
a unit–root–restricted model ${\cal M}_0$ and an unrestricted “stationary” model ${\cal M}_1$,
with corresponding information criteria $IC_{n,0}(\hat{z})$ and $IC_{n,1}(\hat{z})$.
We select ${\cal M}_0$ if $IC_{n,0}(\hat{z})<IC_{n,1}(\hat{z})$ and ${\cal M}_1$ otherwise. With $\Delta IC_n(\hat z) \coloneqq
IC_{n,0}(\hat z) - IC_{n,1}(\hat z)$, $\Delta IC_n(\hat z)<0$ corresponds to selecting the restricted ADF model ${\cal M}_0$
while $\Delta IC_n(\hat z)>0$ corresponds to selecting
the unrestricted model ${\cal M}_1$.
We interpret $\Delta IC_n(\hat z)>0$ as evidence in favour of sparse cointegration
and $\Delta IC_n(\hat z)<0$ as evidence in favour of a spurious regression. The following result shows that this selection rule is asymptotically consistent
under very mild conditions on the time–series behaviour of $\{\hat z_t\}$. \\

\noindent
\emph{Proposition 3. (Residual-Based Cointegration Detection).} \emph{Suppose that $y_{t}$ and ${\bm x}_{t}$ satisfy the framework of Section~2 with $|\rho|\le 1$.
Let $\hat z_t$ be the Stage~1 residuals constructed from the adaptive lasso estimator in (\ref{eq:eq7}).
Assume that the penalty sequence $c_n$ used in \eqref{eq:eq12}–\eqref{eq:eq13} satisfies
$c_n\to\infty$ and $c_n/n\to 0$.
Then, regardless of the cointegration structure among the regressors ${\bm x}_t$,
as $n\to\infty$, 
(i) If $|\rho|<1$, then $P(\Delta IC_n(\hat{z})>0) \to 1$, so the information criterion selects the unrestricted ADF model ${\cal M}_{1}$ and we correctly conclude in favour of sparse cointegration. 
(ii) If $\rho=1$, then $P(\Delta IC_n(\hat{z})<0) \to 1$ so the information criterion selects the unit-root restricted model ${\cal M}_{0}$ and we correctly conclude that the regression is spurious.}\\

Proposition~3 establishes that the residual–based procedure asymptotically selects 
the stationary ADF model when the true errors are stationary and the unit–root–restricted
ADF model when the true errors follow a random walk.
Crucial to the validity of Proposition 3 is the requirement that the $\hat{z}_{t}$'s are prediction consistent under cointegration, and under spurious regression they preserve 
the unit-root behaviour of $z_{t}$ (see Lemma C1 in the appendix). 

A criterion such as the BIC with $c_n=\ln n$ satisfies both $c_n\to\infty$ and
$c_n/n\to 0$ and is therefore model–selection consistent for the problem of discriminating
between ${\cal M}_0$ and ${\cal M}_1$ on the basis of the behaviour of $\hat z_t$.
In practice,
however, different choices of $c_n$ entail different finite–sample trade–offs between
detecting I(0) behaviour and guarding against spurious rejections under I(1). 

\paragraph{Remark 4.}
Proposition~3 imposes no restrictions on the internal cointegration of the regressors $\bm x_t$. The logic relies on the properties of the subspace spanned by $\bm x_t$ rather than the uniqueness of the parameter vector estimated by the adaptive lasso. Under cointegration ($|\rho|<1$), the adaptive lasso maintains prediction consistency even if $\bm x_t$ is rank deficient, identifying a linear combination that eliminates the trend so that $\hat z_t$ is $I(0)$. Conversely, under spurious regression ($\rho=1$), $y_t$ contains a stochastic trend asymptotically independent of $\bm x_t$. Consequently, no linear combination of the regressors can eliminate this trend, ensuring that $\hat z_t$ remains $I(1)$ regardless of the rank or endogeneity of $\bm x_t$. 

\medskip

\noindent 
In the context of Proposition 3, it is also important to emphasise that our main result is driven by the relative orders of magnitude
of the residual sums of squares under ${\cal M}_0$ and ${\cal M}_1$, rather than by the detailed
limiting distribution of any unit–root test statistic.
The requirement
\[
IC_{n,0}(\hat{z}) > IC_{n,1}(\hat{z})
\quad \Longleftrightarrow \quad
n \ln \Bigl(\frac{\overline{\sigma}^2_{0}}{\overline{\sigma}^2_{1}}\Bigr) > c_n
\]
resembles a likelihood–ratio type rejection rule, but our analysis relies only on the fact
that $n \ln (\overline{\sigma}^2_{0}/\overline{\sigma}^2_{1})$ is $O_p(1)$ under the relevant
model. No nuisance–parameter–free limit distribution is needed.
In particular, the generated–regressor aspect of $\hat z_t$ does not affect the asymptotic
selection rule. Replacing $z_t$ by $\hat z_t$ perturbs both residual variances
$\overline{\sigma}^2_{0}$ and $\overline{\sigma}^2_{1}$ by $o_p(1)$, so that the sign of
$\Delta IC_{n}(\hat{z})$ is unchanged asymptotically.
This, however, does not mean that any $c_n$ satisfying $c_n\to\infty$ and $c_n/n\to 0$
will provide accurate decision frequencies in finite samples.
From Proposition~2, for instance, we can infer that the adaptive lasso
will select a large number of I(1) predictors when the true model is spurious (no cointegration).
The resulting residuals $\hat z_t$, obtained using a large number of estimated slope parameters
may require substantial sample sizes in order to mimic an I(1) process accurately, and the finite–sample
behaviour of $n \ln (\overline{\sigma}^2_{0}/\overline{\sigma}^2_{1})$ can be sensitive to
over–selection in the first stage.

\section{Implementation}

\noindent
\textbf{\textit{Penalty term in the adaptive lasso estimation (stage 1)}} 
\medskip

The implementation of the adaptive lasso optimisation programme in \eqref{eq:eq7} requires choosing the penalty parameter $\lambda_{n}$ and setting the adaptive weights $w_{j}$. As is common in the model selection literature, the theoretical requirements on $\lambda_{n}$ leave a multitude of valid choices in practice. In finite samples, $\lambda_n$ is treated as a tuning parameter and selected in a data-driven way.
In the context of unit-root autoregressions, Kock (2016) proposes to select $\lambda_{n}$ via a BIC criterion. This entails estimating the model via the adaptive lasso across a grid of $\lambda_{n}$ values and picking the one that minimises BIC. We adopt the same principle. Letting
\[
RSS(\lambda_{n})
= \sum_{t=1}^{n}\bigl(y_{t}-\hat{\beta}_{0}^{AL}(\lambda_n)-{\bm x}_{t}'\hat{\bm \beta}^{AL}(\lambda_n)\bigr)^{2}
\]
denote the residual sum of squares obtained using a given $\lambda_{n}$, we define
\begin{align}
	\mathrm{BIC}(\lambda_{n}) 
	& = n \ln \Bigl(\frac{\mathrm{RSS}(\lambda_{n})}{n}\Bigr)+ k(\lambda_{n}) \,\ln n,
	\label{eq:eq14}
\end{align}
where $k(\lambda_{n})$ is the number of nonzero coefficients in $\hat{\bm \beta}^{AL}(\lambda_{n})$. The data-based penalty used in our adaptive lasso estimation is then
\begin{align}
	\hat{\lambda}_{n}^{BIC} 
	& = \arg \min_{\lambda \in \Lambda}
	\Bigl\{
	n \ln \Bigl(\frac{\mathrm{RSS}(\lambda)}{n}\Bigr)+ k(\lambda) \,\ln n
	\Bigr\},
	\label{eq:eq15}
\end{align}
where $\Lambda$ is a finite grid of candidate values. In practice, it is convenient to base $\Lambda$ on the usual lasso path: one can first compute the largest penalty $\lambda_{\max}$ for which all slope coefficients are shrunk to zero, and then construct a decreasing grid from a small fraction of $\lambda_{\max}$ up to $\lambda_{\max}$ on a logarithmic scale. In our Monte Carlo experiments we implemented this idea using a grid of 20 logarithmically spaced values spanning a wide range of penalties. Although the BIC-selected $\hat\lambda_n^{BIC}$ is random, the theoretical conditions on $\lambda_n$ ensure that, asymptotically, the admissible range of penalties is broad, and our simulations indicate that BIC tuning produces values of $\hat\lambda_n^{BIC}$ consistent with the behaviour predicted by Propositions~1 and~2.

\medskip
\noindent
\textbf{\textit{Adaptive lasso weights (stage 1)}}
\medskip

Regarding the adaptive weights $w_j$, we follow the standard practice in the adaptive lasso literature (Zou (2006)) and base them on a preliminary estimator $\widetilde{\beta}_j$ of the slope coefficients in \eqref{eq:eq1}, typically the OLS estimator. Concretely, we set $w_{j}=1/|\widetilde{\beta}_{j}^{ols}|^{\gamma}$, $\gamma>0$.
Intuitively, coefficients with larger preliminary estimates receive smaller penalties, while coefficients whose preliminary estimates are close to zero are penalised more heavily. This encourages the adaptive lasso to retain truly active regressors and to shrink irrelevant ones exactly to zero. 
\noindent
The exponent $\gamma$ controls the strength of the adaptation. In applied work
it is most commonly set to $\gamma=1$ or $\gamma=2$. A larger value of $\gamma$ induces more aggressive
penalisation of coefficients with small preliminary estimates, allowing for a
slower growth rate of $\lambda_n$ to achieve model-selection consistency; a smaller $\gamma$ implies milder adaptation and thus requires a faster growth
rate of $\lambda_n$. 

\medskip
\noindent
\textbf{\textit{Optimisation algorithm for the adaptive lasso problem (stage 1)}} 
\medskip

The optimisation problem in \eqref{eq:eq7} is convex, and in practice we compute $(\hat{\beta}_0^{AL},\hat{\bm \beta}^{AL})$ using a standard coordinate-descent implementation of the lasso. Concretely, after computing the initial OLS-based weights $w_{j}$, we rescale the regressors by forming $\widetilde{X}=X \,\mathrm{diag}(w^{-1})$ and fit a plain lasso to $(\widetilde{X},y)$ over the grid $\Lambda$. The resulting lasso estimates, say $\hat{\bm \beta}^{plain}(\lambda)$, are then mapped back to obtain the adaptive lasso counterparts $\hat{\bm\beta}^{AL}(\lambda) = \mathrm{diag}(w)\,\hat{\bm \beta}^{plain}(\lambda)$. This reparametrisation is standard and allows us to exploit existing, numerically stable lasso solvers without having to code a bespoke adaptive lasso routine. 

\medskip
\noindent
\textbf{\textit{Treatment of $\hat{z}$ in the detection of sparse cointegration (stage 2)}} 
\medskip

Our two-step cointegration detection procedure hinges on examining the adaptive-lasso residual series $\{\hat{z}_{t}\}$ via the ADF-type regressions in \eqref{eq:eq10}–\eqref{eq:eq11}. Proposition~2 shows that under no cointegration (spurious regression) the adaptive lasso tends to select nearly all $p$ predictors as active in large samples. Hence for magnitudes of $\rho$ at or near 1, the residual sequence $\hat{z}_{t}$ is likely formed from a highly over-parameterised fit, with many small coefficients. As a result, its sampling variability may deviate from that of a genuine I(1) process in finite samples, possibly yielding a behaviour that wrongly mimics that of an I(0) process. This may make it harder for the BIC–type criterion in Stage~2 to correctly favour the unit–root–restricted ADF model. For this reason, in our empirical implementation we apply a simple capping rule based on which the original residuals are re-estimated. This is purely a finite-sample device as none of our theoretical results need to invoke such a cap under $n\rightarrow \infty$ and $p$ fixed. 
The specifics of the proposed mechanism are motivated and discussed in Section~5 below. 

\medskip
\noindent
\textbf{\textit{Penalty sequence and lag length in the $\hat{z}$ based model-selection procedure (stage 2)}} 
\medskip

A classical BIC type penalty $c_{n}=\ln n$ in \eqref{eq:eq12}–\eqref{eq:eq13} may be too weak unless an unrealistically large sample size is available. Once $\hat{z}_{t}$ is overfitted, adding the single ``lagged-level'' regressor $\hat{z}_{t-1}$ in ${\cal M}_{1}$ can give such a sharp drop in residual variance that BIC's extra penalty term $(\ln n)/n$ may not be enough to compensate for it. As a result, the procedure implemented with a BIC penalty may frequently produce spurious findings of stationarity even when the true $z_{t}$ is I(1). In our empirical implementation we specify the penalty as $c_{n}=c \ln n$, $c>0$,  which clearly satisfies the requirements in Proposition 3 while also allowing us to fine-tune outcomes by experimenting across multiple magnitudes of $c$. 

In addition to the above considerations, our two-step cointegration detection procedure requires fitting ADF-type regressions as in \eqref{eq:eq10}–\eqref{eq:eq11}, and this raises the issue of lag length determination. In line with most of the literature on lag-length selection in ADF-type regressions, we estimate $k$ via a BIC criterion implemented on the fitted model in \eqref{eq:eq11} using a given upper bound $k_{max}$. Note that in the present context the accuracy of $\hat{k}_{BIC}$ is not fundamental for achieving model selection consistency when comparing models ${\cal M}_{0}$ and ${\cal M}_{1}$. This is because the selection consistency result of Proposition~3 does not require serially uncorrelated in \eqref{eq:eq10}–\eqref{eq:eq11}; rather, it depends on the relative orders of the residual variances under I(0) and I(1) specifications, which are robust to mild serial correlation captured by a finite lag length.

\section{Finite Sample Experiments}

\subsection{Monte Carlo evaluation criteria}

The Monte Carlo study is designed to complement the asymptotic results in Propositions~1–3 and to document their finite–sample behaviour under a range of DGPs. In particular, we aim to address three questions:
(i) How well does the adaptive lasso estimate $S$ and the
	corresponding slope coefficients when the conditions of Proposition~1
	hold, and how does its performance deteriorate when the regressors
	exhibit more complicated cointegration patterns (Scenarios labelled as C0–C3 in the DGP design below)?
	(ii) In the spurious–regression case $\rho=1$ of Proposition~2, to what extent
	does the adaptive lasso behave as predicted, and how sensitive
	is this behaviour to cointegration among the regressors?
	(iii) For Proposition~3, does the residual sequence $\{\hat z_t\}$ produced by Stage~1 behave as an $I(0)$ or $I(1)$ process in accordance with the true
	error $z_t$, is $\hat z_t$ a good proxy for $z_t$, and how accurately does the Stage~2 ADF–based model–selection rule discriminate between	$|\rho|<1$ and $\rho=1$?

To answer these questions, we collect three groups of finite–sample metrics from each Monte Carlo design.

\paragraph{Model–selection performance (stage 1).}
For each replication we record the selected set $\widehat S$, and summarise performance across replications through the following metrics.
\emph{Selection error rates:} the false positive rate (FPR) on $S^c$,
	the false negative rate (FNR) on $S$, and the false discovery rate (FDR)
	summarising the proportion of spurious inclusions among selected
	predictors. \emph{Correct set:} the proportion of replications in
	which the adaptive lasso recovers the true active set exactly,
	$\widehat S = S$. \emph{Selected model size:} the average cardinality
	$|\widehat S|$, which reveals whether the method selects parsimonious
	models under $|\rho|<1$ and inflates towards the full support in the
	spurious case $\rho=1$, as suggested by Proposition~2.

\paragraph{Residual behaviour and approximation of $z_t$'s behaviour with $\hat z_t$ (stage 2).}
To support the residual–based arguments underlying Proposition~3, we examine
the behaviour of the true errors $\{z_t\}$ and the adaptive–Lasso residuals
$\{\hat z_t\}$ side by side. For each replication we compute:
the mean squared difference
	$n^{-1}\sum_{t=1}^n (\hat z_t - z_t)^2$, the sample correlation between
	$\{\hat z_t\}$ and $\{z_t\}$, and the ratio of sample variances
	$\widehat{\mathrm{Var}}(\hat z_t)/\widehat{\mathrm{Var}}(z_t)$.
	These quantities quantify how well $\hat z_t$ proxies the latent error
	$z_t$ and thus how severe the generated–regressor effect is in practice.

\paragraph{Residual–based classification between $|\rho|<1$ and $\rho=1$ (stage 2).}
Finally, we assess the performance of the Stage~2 ADF regression-based model–selection
procedure of Proposition~3. For each replication we compute the information
criteria $IC_{n,0}(\cdot)$ and $IC_{n,1}(\cdot)$ for the unit–root–restricted
and unrestricted ADF specifications, respectively, both when they are fitted
to the true errors $z_t$ (oracle benchmark) and when they are fitted to the
estimated residuals $\hat z_t$. We then record the \emph{classification frequencies based on $\hat z_t$} i.e. the proportions
	of replications in which the rule based on
	$\Delta IC_n(\hat z)=IC_{n,0}(\hat z)-IC_{n,1}(\hat z)$ selects the
	stationary specification (interpreted as $|\rho|<1$) or the
	unit–root–restricted specification (interpreted as $\rho=1$), and the
\emph{oracle classification frequencies} i.e. the analogous frequencies
	when the same decision rule is applied to the unobserved true errors
	$z_t$, using $\Delta IC_n(z)$.

\subsection{DGP design and parameterisations}

We consider the single–equation regression
\begin{equation}
	y_t \;=\; \beta_0 + \sum_{j=1}^p \beta_j x_{j,t} + z_t,
	\qquad t=1,\dots,n,
	\label{eq:eq16}
\end{equation}
with $p\in\{10,50,100\}$ and a sparse coefficient vector
on $S=\{1,\dots,5\}$, so that $s\coloneqq |S|=5$ and $|S^c|=p-5$. The five active covariates are taken as the first five of the $x_{j,t}$'s, and we experiment with two signal strengths on the active set, namely $\beta_{S}^{(\text{weak})} = (0.25,\,0.25,\,0.25,\,0.25,\,0.25)'$ and $\beta_{S}^{(\text{strong})} = (1,\,0.5,\,1.5,\,0.8,\,1)'$. The disturbance $z_t$ follows an AR(1) process
\begin{equation}
	z_t = \rho z_{t-1} + e_t,
	\label{eq:eq17}
\end{equation}
with $\rho\in\{0,0.5,0.8, 0.9, 1\}$ and where $e_{t}$ is a stationary random shock (denoted $u_t$ in Section~2); we relabel it here for notational convenience. When $|\rho|<1$, $z_t$ is
stationary and \eqref{eq:eq16} is a cointegrating regression (sparse cointegration).
When $\rho=1$, $z_t$ is a random walk and \eqref{eq:eq16} is a spurious regression. 

Next, we specify the dynamics of the regressors $x_{j,t}$. Although our key objective is to illustrate our theoretical results in Propositions 1-3
we also wish to document what happens when the restrictions related to the potential presence of cointegration among the right hand side covariates are violated. For this 
purpose we design the dynamics of the $x_{j,t}$'s in a way that allows for flexible cointegration patterns on the right-hand side. We introduce $K$ 
common I(1) factors $s_t=(s_{1,t},\dots,s_{K,t})'$:
\begin{equation}
	s_t = s_{t-1} + \xi_t,
	\qquad
	\xi_t = (\xi_{1,t},\dots,\xi_{K,t})',
	\label{eq:eq18}
\end{equation}
and a stationary idiosyncratic component $u_t=(u_{1,t},\dots,u_{p,t})'$.
Each regressor is then defined as
\begin{equation}
	x_{j,t} = s _{g(j),t} + u_{j,t},
	\qquad
	j=1,\dots,p,
	\label{eq:eq19}
\end{equation}
where $g:\{1,\dots,p\}\to\{1,\dots,K\}$ is a group–assignment function
that controls the cointegration structure. If two regressors $x_{j,t}$ and
$x_{k,t}$ share the same group index $g(j)=g(k)$, their difference
$x_{j,t}-x_{k,t}=u_{j,t}-u_{k,t}$ is I(0) and they are cointegrated.
If they do not share any factor, no such exact cointegration arises.
In all designs we take $K=p$ and construct different scenarios by changing
the assignment $g(\cdot)$. This ensures that any desired pattern of
cointegration (or absence) can be induced by suitable grouping. 

To capture both cross–sectional correlation among shocks and potential
endogeneity between regressors and the error term $z_t$, we stack all
innovations into the single vector
\[
\eta_t
:=
\begin{pmatrix}
	\xi_t \\
	u_t \\
	e_t
\end{pmatrix}
\in\mathbb{R}^{K+p+1},
\]
and assume
\begin{equation}
	\eta_t \overset{i.i.d.}{\sim} N(0,\Sigma),
	\qquad
	\Sigma_{ij} = 0.5^{|i-j|}.
	\label{eq:eq20}
\end{equation}
This specification induces:
(i) correlation across the I(1) factors via $\text{cov}(\xi_{k,t},\xi_{\ell,t})$,
(ii) correlation across the stationary components via $\text{cov}(u_{j,t},u_{k,t})$,
and (iii) endogeneity through nonzero covariances between the shocks to $z_t$
and the shocks to the regressors, $\text{cov}(e_t,\xi_{k,t})$ and
$\text{cov}(e_t,u_{j,t})$.
All processes are initiated at zero, $s_0=0$, $u_0=0$ and $z_0=0$. Despite its simplicity, the specification
\eqref{eq:eq16}–\eqref{eq:eq20} captures the three key phenomena
we wish to study: I(1) regressors with possibly complicated cointegration
patterns, serial correlation in the error, and endogeneity between
$\{x_{j,t}\}$ and $z_t$. \\

\noindent
\textbf{Right hand side cointegration scenarios}: The function $g(\cdot)$ in \eqref{eq:eq19} allows us to generate several
distinct designs for the cointegration structure of the right–hand side
covariates (recall that the active set is $S=\{1,\dots,5\}$ and the inactive
set is $S^c=\{6,\dots,p\}$).

\begin{itemize}
	\item {{\bf Scenario C0}: No cointegration among regressors.} Here we set $g(j)=j$ for all $j=1,\dots,p$.
	Each regressor loads on its own factor and there are no cointegrating
	relations among the $x_{j,t}$'s. This corresponds to the classical non–cointegrated regressor design; when combined with $\rho=1$
	it recovers the spurious–regression environment of Proposition~2.

	\item {{\bf Scenario C1}: Cointegration only among inactives.} Here actives load on distinct factors, $g(j)=j$ for $j\in S$.
	The inactives are partitioned into groups that share factors, for
	example pairs: $ g(6)=6,\ g(7)=6; g(8)=7,\ g(9)=7$ etc. 
	Thus cointegration is allowed to occur only within the inactive block
	$x_{S^c,t}$, while any linear combination that places nonzero weight on
	at least one active regressor remains I(1). Combined with $|\rho|<1$, this
	design conforms to the environment of Proposition~1.

	\item {{\bf Scenario C2}: Cointegration among actives only.} A subset of the active regressors share a common factor, while all
	inactives load on distinct factors.
	A simple choice is $g(1)=g(2)=1$, $g(3)=3$, $g(4)=4$, $g(5)=5$ for
	the actives, and $g(j)=j$ for $j\in S^c$.
	Then $x_{1,t}$ and $x_{2,t}$ are cointegrated, but the inactives are
	not cointegrated among themselves nor with the actives.
This violates the cointegration assumptions underlying Propositions~1–2
and serves to illustrate the breakdown of model–selection consistency in
Proposition~1 while preserving the validity of the Stage~2 residual–based
classification of Proposition~3.

	\item {{\bf Scenario C3}: Cointegration in both blocks with cross–block relations.}
	Here both actives and inactives participate in cointegrating relations,
	and some factors are shared across $S$ and $S^c$.
	For instance, one may take $g(1)=g(2)=1$ for two actives, and
	$g(6)=1$ for an inactive, so that $x_{1,t}$, $x_{2,t}$, and $x_{6,t}$
	share a common I(1) trend.
	Additional inactives can be grouped as in Scenario~C1 to generate
	further cointegration among $x_{S^c,t}$.
	This represents a flexible right-hand side with cointegration both
	within and across the active and inactive blocks, and is designed to
	assess the residual behaviour underlying Proposition~3.
\end{itemize}

\subsection{A finite-sample capping mechanism for a cleaner $\hat{z}$}

As noted in Section~4, the adaptive lasso's tendency to overfit under spurious regression ($\rho=1$) typically yields residuals that appear stationary in finite samples, particularly when $p$ is large. While Proposition~3 guarantees consistency as $n \to \infty$, the sample size required for $\hat{z}_t$ to mimic an $I(1)$ process in high-dimensional settings can be substantial. To mitigate this finite-sample pathology while adhering to the sparse-cointegration perspective, we introduce a parsimony cap on the number of selected coefficients:$$s_{\mathrm{cap}}
=
\min\bigl\{p,\,
\max\bigl\{0,\ \mathrm{round}(S_{\mathrm{base}} + F_{\log p}\,\ln p)\bigr\}
\bigr\},$$where $S_{\mathrm{base}}$ and $F_{\log p}$ are user-specified constants (e.g., $S_{\mathrm{base}}=4$, $F_{\log p}=1.5$). In practice, we implement this by retaining only the $s_{\mathrm{cap}}$ coefficients with the largest absolute values from the first-stage estimation. This procedure acts as a finite-sample stabilization device akin to cardinality-constrained regression (e.g., Blumenthal and Davies1; Bertsimas et al. 2). Crucially, with $p$ fixed and $s_{\mathrm{cap}} \ge s$, the cap has no impact on the asymptotic validity of our results. Under cointegration ($|\rho|<1$), the capped support continues to contain the true active set with probability approaching one. Under spurious regression ($\rho=1$), trimming the support reduces overfitting without eliminating the unit-root component of the residuals, thereby ensuring the robustness of the Stage~2 decision rule.

\paragraph{Remark 5.} Our Stage~2 procedure relies on $\hat z_t$ inheriting the correct \emph{order of integration} from $z_t$, rather than converging to it in mean-square. Under $|\rho|<1$, Proposition~1 ensures $\hat z_t$ is a consistent $I(0)$ proxy. In the spurious case ($\rho=1$), however, the uncapped adaptive lasso absorbs substantial random-walk variation from $y_t$ (Proposition~2). The capping mechanism addresses this by limiting the variance reduction achievable through spurious correlations, preserving the $I(1)$ signal required for detection without requiring $\hat z_t$ to be a precise reconstruction of $z_t$. In the empirical analysis that follows, we focus on the capped implementation; comparative results for the uncapped version are provided in the Supplementary Appendix.

\subsection{Empirical Outcomes}

\subsubsection{\textbf{Adaptive lasso estimation and model selection consistency (stage1)}} 

\noindent
{\textit{Favourable designs C0 and C1.}} Scenarios C0 (no cointegration among regressors) and C1 (cointegration among inactives only) satisfy the conditions of Proposition~1. Monte Carlo results confirm the expected behaviour when $|\rho|<1$. For $p=10$ and $\rho\in\{0,0.5\}$, the adaptive lasso performs nearly perfectly. At $n=500$, error rates are negligible (e.g., in C0 with $\rho=0$, $\Pr(\widehat S=S)\approx 0.98$). As $n$ increases to $1000$, both FPR and FNR essentially vanish. As dimensionality increases to $p=50$ and $p=100$, selection becomes more difficult but the pattern holds. While the exact support recovery probability $\Pr(\widehat S=S)$ declines (e.g., to $\approx 0.35$ for $p=100$), the FNR remains extremely low. Crucially, the active covariates are almost always included, ensuring reliable Stage~2 residuals. Scenario C1 behaves similarly to C0, with slightly higher difficulty due to cointegration among inactives, but again, truly active regressors are rarely dropped. As persistence increases ($\rho=0.8, 0.9$), the procedure begins to reflect the influence of near-unit-root errors, with slightly rising FPRs and model sizes. This is partly driven by the capping mechanism, which mitigates overfitting as $\rho \to 1$.

\noindent
{\textit{Scenario C2: cointegration among actives.}}
Scenario C2 violates the identification assumption of Proposition~1. As expected, dropping active regressors becomes common. For instance, with $\rho=0$ and $p=10$, FNR rises to $\approx 0.14$, and $\Pr(\widehat S=S)$ drops to $\approx 0.30$.
In high dimensions ($p=50, 100$), $\Pr(\widehat S=S)$ is effectively zero. The adaptive lasso typically selects a sparse model including only a subset of the cointegrated actives, reflecting the multiplicity of sparse representations available when regressors share stochastic trends. As $\rho$ increases, performance further deteriorates, with FDR exceeding $0.6$–$0.7$ near the unit root boundary.

\begin{center}
{\bf Table 1 about here}
\end{center}

\noindent
{\textit{Scenario C3: cointegration within and across $S$ and $S^c$.}} Scenario C3 allows cointegration both
within the active set and across the active and inactive blocks.
The Monte Carlo results confirm that (capped) adaptive lasso
struggles in this environment even when $\rho$ is small.
For $\rho=0$, $p=10$, $n=500$ we observe large error rates
(FPR $\approx 0.21$, FNR $\approx 0.21$) and a Correct–Set probability
near zero.  

\noindent
{\textit{Behaviour as $\rho\to 1$ and Proposition~2.}} 
Table 1 illustrates the overfitting predicted by Proposition~2. Under $\rho=1$, the selected model size saturates near the imposed upper bound $s_{cap}$ (e.g., 7 for $p=50$).
Table 2 further corroborates this by reporting average selection probabilities. Under cointegration ($\rho \le 0.5$), the probability of selecting actives $\overline{\pi}_{A} \approx 1$ while for inactives $\overline{\pi}_{I} \approx 0$. Under spurious regression ($\rho=1$), $\overline{\pi}_{A}$ drops and $\overline{\pi}_{I}$ increases sharply (e.g., to $\approx 0.47$), confirming the tendency to select spurious predictors. 

\begin{center}
{\bf Table 2 about here}
\end{center}

\paragraph{Remark 6.} The counterpart to Table 1 under no capping imposed on the adapted lasso is Table S1 in the supplementary appendix. We note that key differences only kick in as $\rho$ approaches the unit-root boundary (as expected). Under $p=100$ and $\rho=0.9$ for instance, the capped lasso hovers around $\hat{S}\approx 8$ while without capping $\hat{S}\approx 24$. Such overfitting almost certainly distorts the stochastic properties of $\hat{z}_{t}$ in finite samples as illustrated in Table S4 which presents decision frequencies for sparse cointegration detection (see Supplementary Appendix).

\subsubsection{\textbf{Residual analysis: comparing $\hat{z}$ with $z$}}
	
Table~3 reports Monte Carlo diagnostics for the Stage 2 residuals $\hat z_t$
under the capped implementation, for all scenarios
$C0$--$C3$.  

\textit{Low and moderate persistence ($\rho=0, 0.5$).} When $z_t$ is stationary, $\hat z_t$ tracks the true error extremely well. Correlations are near unity ($0.94$--$1.00$) and variance ratios are close to one. This confirms that when Stage~1 consistently identifies the active regressors, Stage 2 delivers residuals that converge to the true error process.
\begin{center}
	{\bf Table 3 about here}
\end{center}

\textit{Role of cointegration in $x_t$.} Comparing scenarios $C0$--$C3$ shows that the internal cointegration structure of $x_t$ has only a second-order impact on residual diagnostics. Differences in MSE and correlation are marginal and shrink with $n$. This supports Proposition~3. The limiting behaviour of Stage~2 residuals is driven by $z_t$ and is largely invariant to regressor cointegration.

\textit{High persistence and Unit Roots ($\rho \ge 0.8$).} As $\rho \to 1$, approximating $z_t$ becomes difficult. For $\rho \in \{0.8, 0.9\}$, the variance ratio drops below one, indicating "shrinkage" of the residuals. Crucially, the capping mechanism mitigates this: for $p=100, \rho=0.9$, the capped version preserves $\approx 70\%$ of the variance versus $50\%$ for the uncapped version. When $z_t$ is exactly $I(1)$ ($\rho=1$), MSEs become large and variance ratios drop, as expected. However, the capped procedure retains a non-trivial fraction of the unit-root variation, preventing the over-aggressive removal of the trend observed in the uncapped case.

\subsubsection{\textbf{\textit{Sparse cointegration detection (stage2)}}}

Table~4 reports the performance of the Stage~2 ADF model-selection rule.

\textit{Interior stationary regime ($\rho \le 0.8$).} For $|\rho|<1$ and away from the boundary, the procedure performs nearly perfectly. Across all scenarios and dimensions, both the oracle (using true $z_t$) and residual-based rules select the stationary model with probability $\approx 1$. This confirms that Stage~2 is robust to the presence or absence of regressor cointegration.

\textit{Boundary case ($\rho=0.9$).} At $n=500$, even the oracle criterion tends to misclassify $\rho=0.9$ as a unit root ($\approx 80\%$ of replications). Interestingly, the residual-based rule (with capping) is more favourable to stationarity, correcting this bias in higher dimensions. At $n=1000$, both oracle and residual-based decisions align with the asymptotic prediction of stationarity.

\begin{center}{\bf Table 4 about here}\end{center}

\textit{Spurious regression regime ($\rho=1$).} This is central for Proposition~2. The oracle correctly selects the unit-root model with probability $\approx 1$. The residual-based performance depends on $p$ and capping. For low dimension ($p=10$), the rule performs well ($\approx 97$--$99\%$ accuracy). For high dimension ($p=50, 100$), the parsimony cap is essential. The capped rule selects the unit-root model in $80$--$89\%$ of replications (compared to far lower rates for the uncapped version, see Supplementary Appendix). Although strictly lower than the oracle, the cap effectively restores the correct qualitative behaviour, allowing the information criterion to recognise the absence of cointegration in the vast majority of samples.

\textit{Robustness.} The results in Table 4 show that the cointegration structure of the regressors ($C0$--$C3$) has virtually no impact on the decision frequencies. The procedure correctly identifies spurious regressions and cointegrating relationships alike, provided Stage~1 overfitting is controlled via capping.

\section{Summary and Conclusions}

We have proposed a two-step procedure for detecting sparse cointegration in high-dimensional settings. The first stage employs the adaptive lasso to identify a subset of covariates and estimate the long-run relationship. We showed that this estimator achieves model selection consistency in classical settings and retains a sure-screening property even when the regressors exhibit internal cointegration.

The second stage reframes the unit-root testing problem as a model selection procedure. By comparing a unit-root-restricted ADF specification against an unrestricted one via an information criterion, we avoid the non-standard, nuisance-parameter-dependent limiting distributions associated with residual-based tests. Our asymptotic analysis confirms that this decision rule is consistent provided the first-stage residuals inherit the correct order of integration.

A crucial theoretical insight is the adaptive lasso's tendency to over-select predictors under spurious regression ($\rho=1$). This overfitting yields residuals that appear artificially stationary in finite samples. To address this, we introduced a finite-sample parsimony cap and a robust penalty term. While not required for the asymptotic theory, these adjustments are important for preventing spurious findings of cointegration when $p$ is large. Monte Carlo experiments confirm that, away from the unit-root boundary, the procedure mimics an oracle observing the true errors. It maintains high power and controlled size across diverse correlation structures. In spurious regression scenarios, the capped implementation effectively limits overfitting, preserving the unit-root signal in the residuals where the uncapped version fails.

Future work may extend this framework to multiple cointegrating relationships, develop a complete post-selection inference theory, and explore the properties of residual-based decision rules under increasing-dimension asymptotics where $p$ grows with $n$.

\begin{table}[htbp]
	\centering
	\tabcolsep3pt
	\caption{Adaptive lasso (capped) model selection}
	\scalebox{0.8}{\begin{tabular}{rllccccccccccccccc}
			\multicolumn{1}{l}{Capped} &       &       & \multicolumn{5}{c}{$p=10$}            & \multicolumn{5}{c}{$p=50$}            & \multicolumn{5}{c}{$p=100$} \\ \hline
			\multicolumn{1}{c}{scenario} & $\rho$ & \multicolumn{1}{c}{n} & FPR   & FNR   & FDR   & Size  & Correct & FPR   & FNR   & FDR   & Size  & Correct & FPR   & FNR   & FDR   & Size  & Correct \\ \hline
			\multicolumn{1}{l}{C0} & 0.0   & 500   & 0.003 & 0.004 & 0.003 & 4.995 & 0.980 & 0.021 & 0.018 & 0.152 & 5.868 & 0.257 & 0.021 & 0.045 & 0.280 & 6.786 & 0.071 \\
			&       & 1000  & 0.000 & 0.000 & 0.000 & 5.000 & 1.000 & 0.005 & 0.000 & 0.039 & 5.235 & 0.766 & 0.007 & 0.000 & 0.113 & 5.690 & 0.349 \\
			& 0.5   & 500   & 0.017 & 0.030 & 0.017 & 4.934 & 0.859 & 0.043 & 0.085 & 0.288 & 6.522 & 0.072 & 0.034 & 0.114 & 0.417 & 7.693 & 0.012 \\
			&       & 1000  & 0.000 & 0.000 & 0.000 & 5.000 & 1.000 & 0.013 & 0.002 & 0.090 & 5.566 & 0.544 & 0.015 & 0.005 & 0.199 & 6.368 & 0.212 \\
			& 0.8   & 500   & 0.168 & 0.210 & 0.171 & 4.788 & 0.219 & 0.079 & 0.313 & 0.508 & 6.982 & 0.000 & 0.049 & 0.331 & 0.581 & 7.993 & 0.000 \\
			&       & 1000  & 0.030 & 0.040 & 0.030 & 4.948 & 0.816 & 0.057 & 0.118 & 0.366 & 6.963 & 0.004 & 0.039 & 0.138 & 0.460 & 7.980 & 0.002 \\
			& 0.9   & 500   & 0.314 & 0.348 & 0.321 & 4.834 & 0.028 & 0.098 & 0.479 & 0.627 & 6.994 & 0.000 & 0.057 & 0.478 & 0.674 & 7.999 & 0.000 \\
			&       & 1000  & 0.170 & 0.190 & 0.172 & 4.897 & 0.277 & 0.082 & 0.343 & 0.530 & 6.997 & 0.000 & 0.050 & 0.342 & 0.589 & 8.000 & 0.000 \\ \hline
			& 1.0   & 500   & 0.469 & 0.478 & 0.473 & 4.954 & 0.009 & 0.125 & 0.723 & 0.802 & 7.000 & 0.000 & 0.069 & 0.709 & 0.818 & 8.000 & 0.000 \\
			&       & 1000  & 0.485 & 0.488 & 0.486 & 4.981 & 0.005 & 0.125 & 0.726 & 0.804 & 7.000 & 0.000 & 0.070 & 0.729 & 0.830 & 8.000 & 0.000 \\ \hline\hline
			\multicolumn{1}{l}{C1} & 0.0   & 500   & 0.007 & 0.008 & 0.007 & 4.998 & 0.963 & 0.031 & 0.017 & 0.206 & 6.295 & 0.181 & 0.028 & 0.038 & 0.341 & 7.433 & 0.039 \\
			&       & 1000  & 0.000 & 0.000 & 0.000 & 5.000 & 0.999 & 0.013 & 0.000 & 0.096 & 5.601 & 0.488 & 0.010 & 0.000 & 0.137 & 5.908 & 0.357 \\
			& 0.5   & 500   & 0.039 & 0.042 & 0.039 & 4.982 & 0.810 & 0.048 & 0.085 & 0.316 & 6.755 & 0.030 & 0.037 & 0.132 & 0.444 & 7.857 & 0.004 \\
			&       & 1000  & 0.002 & 0.002 & 0.002 & 5.000 & 0.992 & 0.025 & 0.002 & 0.172 & 6.133 & 0.244 & 0.020 & 0.004 & 0.255 & 6.865 & 0.122 \\
			& 0.8   & 500   & 0.173 & 0.205 & 0.176 & 4.839 & 0.208 & 0.078 & 0.301 & 0.500 & 6.993 & 0.000 & 0.050 & 0.349 & 0.593 & 7.993 & 0.000 \\
			&       & 1000  & 0.051 & 0.054 & 0.051 & 4.983 & 0.752 & 0.054 & 0.100 & 0.351 & 6.950 & 0.004 & 0.039 & 0.138 & 0.460 & 7.983 & 0.002 \\
			& 0.9   & 500   & 0.272 & 0.321 & 0.282 & 4.751 & 0.036 & 0.095 & 0.459 & 0.613 & 6.995 & 0.000 & 0.057 & 0.486 & 0.679 & 7.996 & 0.000 \\
			&       & 1000  & 0.175 & 0.188 & 0.176 & 4.935 & 0.272 & 0.078 & 0.303 & 0.501 & 6.993 & 0.000 & 0.049 & 0.338 & 0.586 & 8.000 & 0.000 \\ \hline
			& 1.0   & 500   & 0.364 & 0.385 & 0.372 & 4.897 & 0.020 & 0.117 & 0.651 & 0.750 & 7.000 & 0.000 & 0.068 & 0.693 & 0.808 & 8.000 & 0.000 \\
			&       & 1000  & 0.374 & 0.383 & 0.376 & 4.953 & 0.014 & 0.115 & 0.638 & 0.742 & 6.999 & 0.000 & 0.068 & 0.686 & 0.804 & 8.000 & 0.000 \\ \hline\hline
			\multicolumn{1}{l}{C2} & 0.0   & 500   & 0.007 & 0.164 & 0.008 & 4.216 & 0.182 & 0.018 & 0.170 & 0.147 & 4.958 & 0.061 & 0.018 & 0.182 & 0.273 & 5.838 & 0.014 \\
			&       & 1000  & 0.000 & 0.140 & 0.000 & 4.299 & 0.299 & 0.004 & 0.142 & 0.035 & 4.475 & 0.219 & 0.006 & 0.140 & 0.115 & 4.919 & 0.114 \\
			& 0.5   & 500   & 0.034 & 0.177 & 0.034 & 4.285 & 0.153 & 0.044 & 0.210 & 0.311 & 5.933 & 0.015 & 0.035 & 0.224 & 0.441 & 7.187 & 0.002 \\
			&       & 1000  & 0.002 & 0.150 & 0.002 & 4.256 & 0.248 & 0.011 & 0.152 & 0.085 & 4.722 & 0.151 & 0.013 & 0.148 & 0.193 & 5.470 & 0.071 \\
			& 0.8   & 500   & 0.185 & 0.278 & 0.193 & 4.536 & 0.049 & 0.084 & 0.367 & 0.545 & 6.964 & 0.000 & 0.051 & 0.373 & 0.607 & 7.985 & 0.000 \\
			&       & 1000  & 0.071 & 0.180 & 0.072 & 4.456 & 0.141 & 0.066 & 0.223 & 0.431 & 6.870 & 0.001 & 0.043 & 0.232 & 0.517 & 7.969 & 0.000 \\
			& 0.9   & 500   & 0.325 & 0.384 & 0.336 & 4.705 & 0.005 & 0.100 & 0.499 & 0.642 & 6.993 & 0.000 & 0.057 & 0.490 & 0.681 & 7.999 & 0.000 \\
			&       & 1000  & 0.220 & 0.275 & 0.226 & 4.728 & 0.036 & 0.087 & 0.386 & 0.561 & 6.999 & 0.000 & 0.052 & 0.385 & 0.615 & 8.000 & 0.000 \\ \hline
			& 1.0   & 500   & 0.521 & 0.533 & 0.527 & 4.942 & 0.000 & 0.126 & 0.737 & 0.812 & 6.997 & 0.000 & 0.069 & 0.718 & 0.824 & 8.000 & 0.000 \\
			&       & 1000  & 0.534 & 0.538 & 0.536 & 4.980 & 0.001 & 0.127 & 0.747 & 0.819 & 7.000 & 0.000 & 0.071 & 0.741 & 0.838 & 8.000 & 0.000 \\ \hline\hline
			\multicolumn{1}{l}{C3} & 0.0   & 500   & 0.208 & 0.213 & 0.208 & 4.976 & 0.039 & 0.053 & 0.161 & 0.355 & 6.569 & 0.002 & 0.035 & 0.177 & 0.435 & 7.403 & 0.000 \\
			&       & 1000  & 0.182 & 0.185 & 0.182 & 4.986 & 0.080 & 0.049 & 0.129 & 0.333 & 6.546 & 0.002 & 0.025 & 0.131 & 0.351 & 6.751 & 0.002 \\
			& 0.5   & 500   & 0.226 & 0.234 & 0.227 & 4.957 & 0.045 & 0.062 & 0.200 & 0.407 & 6.794 & 0.000 & 0.042 & 0.230 & 0.501 & 7.795 & 0.000 \\
			&       & 1000  & 0.185 & 0.190 & 0.185 & 4.977 & 0.069 & 0.053 & 0.145 & 0.354 & 6.668 & 0.003 & 0.032 & 0.147 & 0.406 & 7.302 & 0.000 \\
			& 0.8   & 500   & 0.294 & 0.322 & 0.300 & 4.858 & 0.012 & 0.084 & 0.360 & 0.540 & 6.965 & 0.000 & 0.052 & 0.394 & 0.621 & 7.985 & 0.000 \\
			&       & 1000  & 0.220 & 0.228 & 0.221 & 4.964 & 0.047 & 0.068 & 0.225 & 0.440 & 6.939 & 0.001 & 0.044 & 0.242 & 0.524 & 7.973 & 0.001 \\
			& 0.9   & 500   & 0.336 & 0.397 & 0.351 & 4.695 & 0.004 & 0.097 & 0.479 & 0.627 & 6.984 & 0.000 & 0.058 & 0.510 & 0.694 & 7.998 & 0.000 \\
			&       & 1000  & 0.277 & 0.299 & 0.282 & 4.888 & 0.017 & 0.084 & 0.360 & 0.543 & 6.996 & 0.000 & 0.052 & 0.383 & 0.614 & 8.000 & 0.000 \\ \hline
			& 1.0   & 500   & 0.385 & 0.442 & 0.408 & 4.716 & 0.005 & 0.120 & 0.684 & 0.774 & 6.999 & 0.000 & 0.069 & 0.710 & 0.819 & 8.000 & 0.000 \\
			&       & 1000  & 0.392 & 0.427 & 0.406 & 4.823 & 0.002 & 0.121 & 0.687 & 0.776 & 6.997 & 0.000 & 0.069 & 0.718 & 0.823 & 7.998 & 0.000 \\
	\end{tabular}}
	\label{tab:Tab1_capped}
	\parbox{0.96\textwidth}{\vspace{0.6em}
		\footnotesize\emph{Notes:}
		The table reports Monte-Carlo averages of the false positive rate (FPR), false negative rate (FNR),
		false discovery rate (FDR), average selected model size $\lvert\widehat S\rvert$, and the probability of
		selecting the exact active set (Correct) for the capped adaptive lasso, across sample sizes $n$,
		dimensions $p$, persistence levels $\rho$, and cointegration scenarios $C0$--$C3$.}
\end{table}%

\begin{table}[htbp]
	\centering
	\caption{Selection Probabilities for active and inactive covariates (capped)}
	\scalebox{0.88}{\begin{tabular}{rllcccccc}
			\multicolumn{1}{l}{Capped} &       &       & \multicolumn{2}{c}{$p=10$} & \multicolumn{2}{c}{$p=50$} & \multicolumn{2}{c}{$p=100$} \\ \hline
			\multicolumn{1}{c}{scenario} & \multicolumn{1}{c}{$\rho$} & \multicolumn{1}{c}{n} & $\overline{\pi}_{A}$ & $\overline{\pi}_{I}$ & $\overline{\pi}_{A}$ & $\overline{\pi}_{I}$ & $\overline{\pi}_{A}$ & $\overline{\pi}_{I}$ \\ \hline
			\multicolumn{1}{l}{C0} & 0.0   & 500   & 0.996 & 0.003 & 0.982 & 0.021 & 0.955 & 0.021 \\
			&       & 1000  & 1.000 & 0.000 & 1.000 & 0.005 & 1.000 & 0.007 \\
			& 0.5   & 500   & 0.970 & 0.017 & 0.915 & 0.043 & 0.886 & 0.034 \\
			&       & 1000  & 1.000 & 0.000 & 0.998 & 0.013 & 0.995 & 0.015 \\
			& 0.8   & 500   & 0.790 & 0.168 & 0.687 & 0.079 & 0.669 & 0.049 \\
			&       & 1000  & 0.960 & 0.030 & 0.882 & 0.057 & 0.862 & 0.039 \\
			& 0.9   & 500   & 0.652 & 0.314 & 0.521 & 0.098 & 0.522 & 0.057 \\
			&       & 1000  & 0.810 & 0.170 & 0.657 & 0.082 & 0.658 & 0.050 \\ \hline
			& 1.0   & 500   & 0.522 & 0.469 & 0.277 & 0.125 & 0.291 & 0.069 \\
			&       & 1000  & 0.512 & 0.485 & 0.274 & 0.125 & 0.271 & 0.070 \\ \hline\hline
			\multicolumn{1}{l}{C1} & 0.0   & 500   & 0.992 & 0.007 & 0.983 & 0.031 & 0.962 & 0.028 \\
			&       & 1000  & 1.000 & 0.000 & 1.000 & 0.013 & 1.000 & 0.010 \\
			& 0.5   & 500   & 0.958 & 0.039 & 0.915 & 0.048 & 0.868 & 0.037 \\
			&       & 1000  & 0.998 & 0.002 & 0.998 & 0.025 & 0.996 & 0.020 \\
			& 0.8   & 500   & 0.795 & 0.173 & 0.699 & 0.078 & 0.651 & 0.050 \\
			&       & 1000  & 0.946 & 0.051 & 0.900 & 0.054 & 0.862 & 0.039 \\
			& 0.9   & 500   & 0.679 & 0.272 & 0.541 & 0.095 & 0.514 & 0.057 \\
			&       & 1000  & 0.812 & 0.175 & 0.697 & 0.078 & 0.662 & 0.049 \\ \hline
			& 1.0   & 500   & 0.615 & 0.364 & 0.349 & 0.117 & 0.307 & 0.068 \\
			&       & 1000  & 0.617 & 0.374 & 0.362 & 0.115 & 0.314 & 0.068 \\ \hline\hline
			\multicolumn{1}{l}{C2} & 0.0   & 500   & 0.836 & 0.007 & 0.830 & 0.018 & 0.818 & 0.018 \\
			&       & 1000  & 0.860 & 0.000 & 0.858 & 0.004 & 0.860 & 0.006 \\
			& 0.5   & 500   & 0.823 & 0.034 & 0.790 & 0.044 & 0.776 & 0.035 \\
			&       & 1000  & 0.850 & 0.002 & 0.848 & 0.011 & 0.852 & 0.013 \\
			& 0.8   & 500   & 0.722 & 0.185 & 0.633 & 0.084 & 0.627 & 0.051 \\
			&       & 1000  & 0.820 & 0.071 & 0.777 & 0.066 & 0.768 & 0.043 \\
			& 0.9   & 500   & 0.616 & 0.325 & 0.501 & 0.100 & 0.510 & 0.057 \\
			&       & 1000  & 0.725 & 0.220 & 0.614 & 0.087 & 0.615 & 0.052 \\ \hline
			& 1.0   & 500   & 0.467 & 0.521 & 0.263 & 0.126 & 0.282 & 0.069 \\
			&       & 1000  & 0.462 & 0.534 & 0.253 & 0.127 & 0.259 & 0.071 \\ \hline\hline
			\multicolumn{1}{l}{C3} & 0.0   & 500   & 0.787 & 0.208 & 0.839 & 0.053 & 0.823 & 0.035 \\
			&       & 1000  & 0.815 & 0.182 & 0.871 & 0.049 & 0.869 & 0.025 \\
			& 0.5   & 500   & 0.766 & 0.226 & 0.800 & 0.062 & 0.770 & 0.042 \\
			&       & 1000  & 0.810 & 0.185 & 0.855 & 0.053 & 0.853 & 0.032 \\
			& 0.8   & 500   & 0.678 & 0.294 & 0.640 & 0.084 & 0.606 & 0.052 \\
			&       & 1000  & 0.772 & 0.220 & 0.775 & 0.068 & 0.758 & 0.044 \\
			& 0.9   & 500   & 0.603 & 0.336 & 0.521 & 0.097 & 0.490 & 0.058 \\
			&       & 1000  & 0.701 & 0.277 & 0.640 & 0.084 & 0.617 & 0.052 \\
			& 1.0   & 500   & 0.558 & 0.385 & 0.316 & 0.120 & 0.290 & 0.069 \\
			&       & 1000  & 0.573 & 0.392 & 0.313 & 0.121 & 0.282 & 0.069 \\
	\end{tabular}}
	\label{tab:Tab2_capped}
	\parbox{0.58\textwidth}{\vspace{0.6em}
		\footnotesize\emph{Notes:}
		The table reports the average selection probabilities among actives $j\leq s$, $\overline{\pi}_{A}=\sum_{j\in S}P(\hat{\beta}_{j}\neq 0)/s$ and inactives $\overline{\pi}_{I}=\sum_{j\in S^{c}}P(\hat{\beta}_{j}\neq 0)/s$ under the capped adaptive lasso and across $n$, $p$, $\rho$ and scenarios $C0$--$C3$. }
\end{table}%

\begin{table}[htbp]
	\centering
	\caption{Residual diagnostics for the adaptive lasso (capped)}
\scalebox{0.8}{\begin{tabular}{rllccccccccc}
		\multicolumn{1}{l}{Capped} &       &       & \multicolumn{3}{c}{$p=10$} & \multicolumn{3}{c}{$p=50$} & \multicolumn{3}{c}{$p=100$} \\ \hline
		\multicolumn{1}{c}{scenario} & \multicolumn{1}{c}{$\rho$} & \multicolumn{1}{c}{n} & MSE   & Corr  & VR    & MSE   & Corr  & VR    & MSE   & Corr  & VR \\ \hline
		\multicolumn{1}{l}{C0} & 0     & 500   & 0.12  & 0.99  & 1.01  & 0.24  & 0.97  & 1.01  & 0.34  & 0.96  & 1.02 \\
		&       & 1000  & 0.04  & 1.00  & 1.00  & 0.05  & 0.99  & 1.00  & 0.08  & 0.99  & 1.00 \\
		& 0.5   & 500   & 0.38  & 0.97  & 1.00  & 0.61  & 0.95  & 0.99  & 0.69  & 0.94  & 0.98 \\
		&       & 1000  & 0.13  & 0.99  & 0.99  & 0.17  & 0.99  & 0.99  & 0.22  & 0.98  & 0.99 \\
		& 0.8   & 500   & 1.84  & 0.92  & 0.94  & 2.63  & 0.88  & 0.88  & 2.76  & 0.88  & 0.85 \\
		&       & 1000  & 0.93  & 0.96  & 0.98  & 1.57  & 0.93  & 0.96  & 1.73  & 0.93  & 0.94 \\
		& 0.9   & 500   & 5.13  & 0.89  & 0.84  & 7.40  & 0.82  & 0.73  & 7.65  & 0.82  & 0.69 \\
		&       & 1000  & 3.43  & 0.93  & 0.93  & 5.38  & 0.87  & 0.87  & 5.55  & 0.87  & 0.84 \\ \hline
		& 1     & 500   & 907.93 & 0.49  & 0.30  & 1004.72 & 0.47  & 0.16  & 912.51 & 0.49  & 0.15 \\
		&       & 1000  & 1939.56 & 0.48  & 0.29  & 1900.45 & 0.47  & 0.16  & 2047.58 & 0.47  & 0.14 \\ \hline\hline
		\multicolumn{1}{l}{C1} & 0     & 500   & 0.16  & 0.98  & 1.01  & 0.21  & 0.97  & 1.01  & 0.31  & 0.96  & 1.01 \\
		&       & 1000  & 0.06  & 0.99  & 1.00  & 0.05  & 0.99  & 1.00  & 0.06  & 0.99  & 1.00 \\
		& 0.5   & 500   & 0.46  & 0.96  & 1.01  & 0.60  & 0.95  & 0.99  & 0.74  & 0.93  & 0.98 \\
		&       & 1000  & 0.18  & 0.99  & 1.00  & 0.18  & 0.99  & 0.99  & 0.22  & 0.98  & 0.99 \\
		& 0.8   & 500   & 1.79  & 0.93  & 0.94  & 2.49  & 0.89  & 0.88  & 2.80  & 0.88  & 0.86 \\
		&       & 1000  & 1.08  & 0.96  & 0.98  & 1.44  & 0.94  & 0.95  & 1.71  & 0.93  & 0.94 \\
		& 0.9   & 500   & 4.72  & 0.90  & 0.85  & 7.19  & 0.83  & 0.74  & 7.61  & 0.82  & 0.70 \\
		&       & 1000  & 3.38  & 0.93  & 0.93  & 4.83  & 0.89  & 0.86  & 5.38  & 0.88  & 0.84 \\ \hline
		& 1     & 500   & 902.64 & 0.51  & 0.33  & 1003.05 & 0.47  & 0.18  & 911.14 & 0.49  & 0.16 \\
		&       & 1000  & 1933.75 & 0.49  & 0.31  & 1900.14 & 0.47  & 0.17  & 2051.06 & 0.47  & 0.14 \\ \hline\hline
		\multicolumn{1}{l}{C2} & 0     & 500   & 0.13  & 0.99  & 1.01  & 0.20  & 0.98  & 1.01  & 0.26  & 0.97  & 1.01 \\
		&       & 1000  & 0.08  & 0.99  & 1.01  & 0.09  & 0.99  & 1.01  & 0.10  & 0.99  & 1.00 \\
		& 0.5   & 500   & 0.29  & 0.98  & 0.99  & 0.48  & 0.96  & 0.98  & 0.54  & 0.95  & 0.97 \\
		&       & 1000  & 0.14  & 0.99  & 1.00  & 0.17  & 0.99  & 1.00  & 0.20  & 0.98  & 0.99 \\
		& 0.8   & 500   & 1.42  & 0.94  & 0.94  & 2.35  & 0.90  & 0.86  & 2.53  & 0.89  & 0.84 \\
		&       & 1000  & 0.68  & 0.97  & 0.97  & 1.26  & 0.95  & 0.94  & 1.42  & 0.94  & 0.92 \\
		& 0.9   & 500   & 4.59  & 0.90  & 0.84  & 7.16  & 0.83  & 0.73  & 7.40  & 0.82  & 0.69 \\
		&       & 1000  & 2.79  & 0.94  & 0.92  & 4.76  & 0.89  & 0.85  & 5.03  & 0.88  & 0.83 \\ \hline
		& 1     & 500   & 902.49 & 0.51  & 0.32  & 1003.77 & 0.47  & 0.17  & 912.87 & 0.49  & 0.15 \\
		&       & 1000  & 1936.60 & 0.49  & 0.30  & 1900.21 & 0.47  & 0.17  & 2046.44 & 0.48  & 0.14 \\ \hline\hline
		\multicolumn{1}{l}{C3} & 0     & 500   & 0.40  & 0.95  & 1.04  & 0.28  & 0.97  & 0.96  & 0.33  & 0.96  & 0.97 \\
		&       & 1000  & 0.21  & 0.97  & 1.02  & 0.17  & 0.98  & 0.96  & 0.17  & 0.98  & 0.96 \\
		& 0.5   & 500   & 0.61  & 0.95  & 1.01  & 0.58  & 0.95  & 0.95  & 0.65  & 0.94  & 0.95 \\
		&       & 1000  & 0.33  & 0.97  & 1.01  & 0.27  & 0.98  & 0.96  & 0.28  & 0.98  & 0.96 \\
		& 0.8   & 500   & 1.67  & 0.93  & 0.95  & 2.29  & 0.90  & 0.87  & 2.57  & 0.89  & 0.85 \\
		&       & 1000  & 1.08  & 0.96  & 0.98  & 1.23  & 0.95  & 0.95  & 1.47  & 0.94  & 0.93 \\
		& 0.9   & 500   & 4.22  & 0.91  & 0.87  & 6.75  & 0.84  & 0.74  & 7.41  & 0.82  & 0.70 \\
		&       & 1000  & 2.98  & 0.94  & 0.93  & 4.33  & 0.90  & 0.87  & 4.89  & 0.89  & 0.84 \\ \hline
		& 1     & 500   & 895.87 & 0.54  & 0.39  & 1001.52 & 0.48  & 0.18  & 910.26 & 0.50  & 0.16 \\
		&       & 1000  & 1921.73 & 0.52  & 0.37  & 1907.58 & 0.46  & 0.18  & 2046.50 & 0.48  & 0.15 \\
	\end{tabular}}
	\label{tab:addlabel}
		\parbox{0.72\textwidth}{\vspace{0.6em}
		\footnotesize\emph{Notes:}
		The table reports the mean-squared deviation $\mathrm{MSE}$, the correlation
		$\mathrm{corr}(\widehat z, z)$, and the variance ratio $\mathrm{VR} = \mathrm{var}(\widehat z)/\mathrm{var}(z)$
		between the Stage~2 residuals and the true disturbance $z_t$, for the capped implementation of the adaptive lasso.
}
\end{table}%

\begin{table}[htbp]
	\centering
	\caption{Detection of Sparse Cointegration (capped adaptive lasso residuals)}
\scalebox{0.8}{\begin{tabular}{lllcccccccccccc}
		Capped &       &       & \multicolumn{4}{c}{$p=10$}    & \multicolumn{4}{c}{$p=50$}    & \multicolumn{4}{c}{$p=100$} \\ \hline
		&       &       & \multicolumn{2}{c}{Oracle} & \multicolumn{2}{c}{$\hat{z}$} & \multicolumn{2}{c}{Oracle} & \multicolumn{2}{c}{$\hat{z}$} & \multicolumn{2}{c}{Oracle} & \multicolumn{2}{c}{$\hat{z}$} \\ \hline
		scenario & \multicolumn{1}{c}{$\rho$} & \multicolumn{1}{c}{n} & $I(0)$ & $I(1)$ & $I(0)$ & $I(1)$ & $I(0)$ & $I(1)$ & $I(0)$ & $I(1)$ & $I(0)$ & $I(1)$ & $I(0)$ & $I(1)$ \\ \hline
		C0    & 0.0   & 500   & 1.000 & 0.000 & 1.000 & 0.000 & 1.000 & 0.000 & 1.000 & 0.000 & 1.000 & 0.000 & 1.000 & 0.000 \\
		& 0.0   & 1000  & 1.000 & 0.000 & 1.000 & 0.000 & 1.000 & 0.000 & 1.000 & 0.000 & 1.000 & 0.000 & 1.000 & 0.000 \\
		& 0.5   & 500   & 1.000 & 0.000 & 1.000 & 0.000 & 1.000 & 0.000 & 1.000 & 0.000 & 1.000 & 0.000 & 1.000 & 0.000 \\
		& 0.5   & 1000  & 1.000 & 0.000 & 1.000 & 0.000 & 1.000 & 0.000 & 1.000 & 0.000 & 1.000 & 0.000 & 1.000 & 0.000 \\
		& 0.8   & 500   & 0.998 & 0.002 & 0.998 & 0.002 & 0.998 & 0.002 & 1.000 & 0.000 & 1.000 & 0.000 & 1.000 & 0.000 \\
		& 0.8   & 1000  & 1.000 & 0.000 & 1.000 & 0.000 & 1.000 & 0.000 & 1.000 & 0.000 & 1.000 & 0.000 & 1.000 & 0.000 \\
		& 0.9   & 500   & 0.171 & 0.829 & 0.615 & 0.385 & 0.189 & 0.811 & 0.877 & 0.123 & 0.169 & 0.831 & 0.916 & 0.084 \\
		& 0.9   & 1000  & 0.994 & 0.006 & 0.999 & 0.001 & 0.995 & 0.005 & 1.000 & 0.000 & 0.993 & 0.007 & 1.000 & 0.000 \\ \hline
		& 1.0   & 500   & 0.000 & 1.000 & 0.034 & 0.966 & 0.000 & 1.000 & 0.198 & 0.802 & 0.000 & 1.000 & 0.214 & 0.786 \\
		& 1.0   & 1000  & 0.000 & 1.000 & 0.011 & 0.989 & 0.000 & 1.000 & 0.159 & 0.841 & 0.000 & 1.000 & 0.207 & 0.793 \\ \hline \hline
		C1    & 0.0   & 500   & 1.000 & 0.000 & 1.000 & 0.000 & 1.000 & 0.000 & 1.000 & 0.000 & 1.000 & 0.000 & 1.000 & 0.000 \\
		& 0.0   & 1000  & 1.000 & 0.000 & 1.000 & 0.000 & 1.000 & 0.000 & 1.000 & 0.000 & 1.000 & 0.000 & 1.000 & 0.000 \\
		& 0.5   & 500   & 1.000 & 0.000 & 1.000 & 0.000 & 1.000 & 0.000 & 1.000 & 0.000 & 1.000 & 0.000 & 1.000 & 0.000 \\
		& 0.5   & 1000  & 1.000 & 0.000 & 1.000 & 0.000 & 1.000 & 0.000 & 1.000 & 0.000 & 1.000 & 0.000 & 1.000 & 0.000 \\
		& 0.8   & 500   & 0.998 & 0.002 & 0.998 & 0.002 & 0.998 & 0.002 & 1.000 & 0.000 & 1.000 & 0.000 & 0.999 & 0.001 \\
		& 0.8   & 1000  & 1.000 & 0.000 & 1.000 & 0.000 & 1.000 & 0.000 & 1.000 & 0.000 & 1.000 & 0.000 & 1.000 & 0.000 \\
		& 0.9   & 500   & 0.171 & 0.829 & 0.553 & 0.447 & 0.189 & 0.811 & 0.880 & 0.120 & 0.169 & 0.831 & 0.923 & 0.077 \\
		& 0.9   & 1000  & 0.994 & 0.006 & 0.999 & 0.001 & 0.995 & 0.005 & 1.000 & 0.000 & 0.993 & 0.007 & 0.998 & 0.002 \\ \hline
		& 1.0   & 500   & 0.000 & 1.000 & 0.015 & 0.985 & 0.000 & 1.000 & 0.146 & 0.854 & 0.000 & 1.000 & 0.195 & 0.805 \\
		& 1.0   & 1000  & 0.000 & 1.000 & 0.007 & 0.993 & 0.000 & 1.000 & 0.130 & 0.870 & 0.000 & 1.000 & 0.180 & 0.820 \\ \hline \hline
		C2    & 0.0   & 500   & 1.000 & 0.000 & 1.000 & 0.000 & 1.000 & 0.000 & 1.000 & 0.000 & 1.000 & 0.000 & 1.000 & 0.000 \\
		& 0.0   & 1000  & 1.000 & 0.000 & 1.000 & 0.000 & 1.000 & 0.000 & 1.000 & 0.000 & 1.000 & 0.000 & 1.000 & 0.000 \\
		& 0.5   & 500   & 1.000 & 0.000 & 1.000 & 0.000 & 1.000 & 0.000 & 1.000 & 0.000 & 1.000 & 0.000 & 1.000 & 0.000 \\
		& 0.5   & 1000  & 1.000 & 0.000 & 1.000 & 0.000 & 1.000 & 0.000 & 1.000 & 0.000 & 1.000 & 0.000 & 1.000 & 0.000 \\
		& 0.8   & 500   & 0.998 & 0.002 & 0.999 & 0.001 & 0.998 & 0.002 & 1.000 & 0.000 & 1.000 & 0.000 & 1.000 & 0.000 \\
		& 0.8   & 1000  & 1.000 & 0.000 & 1.000 & 0.000 & 1.000 & 0.000 & 1.000 & 0.000 & 1.000 & 0.000 & 1.000 & 0.000 \\
		& 0.9   & 500   & 0.171 & 0.829 & 0.570 & 0.430 & 0.189 & 0.811 & 0.881 & 0.119 & 0.169 & 0.831 & 0.923 & 0.077 \\
		& 0.9   & 1000  & 0.994 & 0.006 & 0.999 & 0.001 & 0.995 & 0.005 & 1.000 & 0.000 & 0.993 & 0.007 & 1.000 & 0.000 \\ \hline
		& 1.0   & 500   & 0.000 & 1.000 & 0.022 & 0.978 & 0.000 & 1.000 & 0.187 & 0.813 & 0.000 & 1.000 & 0.211 & 0.789 \\
		& 1.0   & 1000  & 0.000 & 1.000 & 0.011 & 0.989 & 0.000 & 1.000 & 0.161 & 0.839 & 0.000 & 1.000 & 0.188 & 0.812 \\\hline \hline
		C3    & 0.0   & 500   & 1.000 & 0.000 & 1.000 & 0.000 & 1.000 & 0.000 & 1.000 & 0.000 & 1.000 & 0.000 & 1.000 & 0.000 \\
		& 0.0   & 1000  & 1.000 & 0.000 & 1.000 & 0.000 & 1.000 & 0.000 & 1.000 & 0.000 & 1.000 & 0.000 & 1.000 & 0.000 \\
		& 0.5   & 500   & 1.000 & 0.000 & 1.000 & 0.000 & 1.000 & 0.000 & 1.000 & 0.000 & 1.000 & 0.000 & 1.000 & 0.000 \\
		& 0.5   & 1000  & 1.000 & 0.000 & 1.000 & 0.000 & 1.000 & 0.000 & 1.000 & 0.000 & 1.000 & 0.000 & 1.000 & 0.000 \\
		& 0.8   & 500   & 0.998 & 0.002 & 0.999 & 0.001 & 0.998 & 0.002 & 1.000 & 0.000 & 1.000 & 0.000 & 1.000 & 0.000 \\
		& 0.8   & 1000  & 1.000 & 0.000 & 1.000 & 0.000 & 1.000 & 0.000 & 1.000 & 0.000 & 1.000 & 0.000 & 1.000 & 0.000 \\
		& 0.9   & 500   & 0.171 & 0.829 & 0.475 & 0.525 & 0.189 & 0.811 & 0.868 & 0.132 & 0.169 & 0.831 & 0.919 & 0.081 \\
		& 0.9   & 1000  & 0.994 & 0.006 & 0.998 & 0.002 & 0.995 & 0.005 & 1.000 & 0.000 & 0.993 & 0.007 & 1.000 & 0.000 \\ \hline
		& 1.0   & 500   & 0.000 & 1.000 & 0.006 & 0.994 & 0.000 & 1.000 & 0.112 & 0.888 & 0.000 & 1.000 & 0.196 & 0.804 \\
		& 1.0   & 1000  & 0.000 & 1.000 & 0.001 & 0.999 & 0.000 & 1.000 & 0.123 & 0.877 & 0.000 & 1.000 & 0.187 & 0.813 \\
	\end{tabular}}
	\label{tab:addlabel}
	\parbox{0.90\textwidth}{\vspace{0.6em}
\footnotesize\emph{Notes:} The table compares the information-criterion decision in Stage~2 based on the true errors (oracle) and on the
capped residuals $\widehat z_t$, reporting the Monte Carlo frequencies of selecting the stationary model $I(0)$
versus the unit-root-restricted model $I(1)$ across all designs.}
\end{table}%

\newpage

\begin{center}
	{\bf APPENDIX: PROOFS}
\end{center}

\noindent
{\bf Notation.} (i) For a vector $\bm v\in \mathds{R}^{p}$, the sign vector is
$sgn(\bm v)=(sgn(v_{1}),\ldots,sgn(v_{p}))'$, where for any $c\in \mathds{R}$,
$sgn(c)=1$ if $c>0$, $sgn(c)=0$ if $c=0$, and $sgn(c)=-1$ if $c<0$.
(ii) The subdifferential set of $\|\cdot\|_{1}$ at ${\bm v}$ is
$\partial \|\bm v\|_{1}\coloneqq (\partial |v_{1}|,\ldots,\partial |v_{p}|)'$,
where $\partial |v_{i}|=\{1\}$ if $v_{i}>0$, $\partial |v_{i}|=[-1,1]$ if
$v_{i}=0$, and $\partial |v_{i}|=\{-1\}$ if $v_{i}<0$. Thus
$\partial|v_{i}|=sgn(v_{i})$ if $v_{i}\neq 0$ and $\partial|v_{i}|=[-1,1]$ if
$v_{i}=0$.\\

\noindent
{\textbf{Unit root / cointegration limits}.}
We work under the operating assumptions stated in Section~2. Under these
assumptions the following large–sample results are standard in the unit–root
and cointegration literature (see, e.g., Phillips and Durlauf (1986)):
\begin{align}
	{\bm A}_{n} \coloneqq \frac{1}{n^{2}} \sum_{t=1}^{n} 
	{\bm x}_{t}{\bm x}_{t}' & \Rightarrow {\bm A} 	\label{eq:eq21} \\	
	{\bm B}_{n} \coloneqq \frac{1}{n}\sum_{t=1}^{n} {\bm x}_{t}z_{t} & \Rightarrow \bm B, \label{eq:eq22} 
\end{align}
for some random stochastically bounded matrices $\bm A,\bm B$. Note that $\bm B$ is bounded only when $z_t$ is $I(0)$. If $z_t$ is $I(1)$, the term in \eqref{eq:eq22} is $O_p(n)$. Also, the limiting matrix ${\bm A}$ may be rank deficient if the ${\bm x}_{t}'s$ are cointegrated themselves. Partitioning ${\bm x}_{t}$ as $({\bm x}_{S,t}',{\bm x}_{S^{c},t}')'$ we refer to the corresponding 
blocks of ${\bm A}$ as ${\bm A}_{S.S}$, ${\bm  A}_{S.S^{c}}$, ${\bm A}_{S^{c}.S}$, and ${\bm A}_{S^{c}.S^{c}}$. The presence of cointegration solely within the inactive regressors would imply a rank deficient ${\bm A}_{S^{c}.S^{c}}$ while ${\bm A}_{S.S}\succ 0$.

\medskip
Our proofs of Propositions 1-2 distinguish between two alternative designs for the p I(1) regressors. We consider two cases: (i) there is no right-hand side cointegration within ${\bm x}_{t}$ itself, that is, $\forall a\neq 0: a'{\bm x}_{t} \sim I(1)$. Equivalently $\sum_{t}{\bm  x}_{t}{\bm x}_{t}'/n^{2}\Rightarrow \bm A\succ 0$. We will refer to this regime as $\textbf{X-NC}$. (ii) Only the inactive right-hand side predictors inside $S^{c}$ may be cointegrated, but in a way that does not involve the actives (i.e., no cointegration within ${\bm x}_{S,t}$, no cointegration between ${\bm x}_{S,t}$ and ${\bm x}_{S^{c},t}$, but ${\bm x}_{S^{c},t}$ may have an internal cointegrating structure). We will refer to this regime that implies a rank deficient ${\bm A}_{S^{c}.S^{c}}$ as \textbf{X-IC}. Note also that these restrictions will not be needed for establishing the validity of Proposition 3. 
Finally, and solely for notational simplicity, all our proofs omit the fitted intercept. \\

\noindent
{\textbf{Preliminary Lemmas}.} We next introduce a series of intermediate results. Lemmas A1 and A2 relate to the asymptotic properties of the initial OLS estimators and their implications for the behaviour of the adaptive lasso weights. Some of these intermediate results follow from existing results in the unit-root and cointegration literature (e.g., Phillips and Durlauf (1986)). We therefore restrict their proofs to their implications for the asymptotic behaviour of the adaptive lasso weights $w_{j}=|\widetilde{\beta}_{j}^{ols}|^{-\gamma}$, which are invoked in the proofs of Propositions 1-2.
Lemma A3 in turn recalls the asymptotic properties of the score terms $\sum_{t} x_{j,t}z_{t}$.  \\

\noindent
\emph{Lemma A1 (Weights under Cointegration $|\rho| < 1$).}
\emph{Suppose $y_{t}$ and ${\bm x}_{t}$ satisfy the framework of Section 2 with $|\rho|<1$ (stationary errors).}
\begin{enumerate}
	\item[(1)] \emph{(Active block).} Assume ${\bm x}_{S,t}$ is not internally cointegrated and not cointegrated with ${\bm x}_{S^{c},t}$. Then $n(\widetilde{\bm \beta}_{S}^{ols}-\bm \beta_{S}^{0})=O_{p}(1)$. Consequently, for each $j \in S$, $\widetilde{\beta}_{j}^{ols}\xrightarrow{p} \beta_{j}^{0} \neq 0$, implying $w_{j}=|\widetilde{\beta}_{j}^{ols}|^{-\gamma}\xrightarrow{p} |\beta_{j}^{0}|^{-\gamma} \in (0,\infty)$.
	\item[(2)] \emph{(Inactives under \textbf{X-NC}).} If in addition (\textbf{X-NC}) holds, then for each $j \in S^{c}$, $n\widetilde{\beta}_{j}^{ols}\Rightarrow L_{j}$, where $L_{j}$ is a non-degenerate random variable with $P(L_{j}=0)=0$. Consequently, $w_{j}=O_{p}(n^{\gamma})$.
\end{enumerate}
 
\medskip
 
\noindent
\emph{Proof of Lemma A1}
 	Part (1) is a standard result for OLS in cointegrating regressions (super-consistency). We focus on Part (2). Under (\textbf{X-NC}), for $j \in S^{c}$, standard asymptotics imply $n(\widetilde{\beta}_{j}^{ols}-\beta_{j}^{0}) \Rightarrow L_j$. Since $\beta_j^0 = 0$, $n\widetilde{\beta}_{j}^{ols} \Rightarrow L_j$. The limit $L_j$ is a functional of Brownian motions possessing a continuous distribution with $P(L_j=0)=0$.
 	To show $w_j = O_p(n^\gamma)$, observe that $w_j n^{-\gamma} = |n \widetilde{\beta}_j^{ols}|^{-\gamma}$. We must show this quantity is stochastically bounded. For any $\epsilon > 0$, we rely on the continuous mapping theorem and the property $P(L_j=0)=0$. Choose $\delta > 0$ such that $P(|L_j| < \delta) < \epsilon/2$. By weak convergence, there exists $N$ such that for all $n \geq N$, $P(|n \widetilde{\beta}_j^{ols}| < \delta) < \epsilon$.
 	Let $M = \delta^{-\gamma}$. Then:
 	\[
 	P(w_j n^{-\gamma} > M) = P(|n \widetilde{\beta}_j^{ols}|^{-\gamma} > \delta^{-\gamma}) = P(|n \widetilde{\beta}_j^{ols}| < \delta) < \epsilon.
 	\]
 	Thus, $w_j n^{-\gamma} = O_p(1)$, or $w_j = O_p(n^\gamma)$.  \qed \\

\medskip
\noindent
\emph{Lemma A2 (OLS under Spurious Regression $\rho=1$).}
\emph{Suppose the conditions of Lemma A1(1) hold but the errors are $I(1)$ (i.e., $\rho=1$, so $y_t$ and ${\bm x}_t$ are not cointegrated). Then:
(1) $\widetilde{\beta}_{j}^{ols}=O_{p}(1)$ and $\widetilde{\beta}_{j}^{ols}\Rightarrow \xi_{j}$, where $\xi_{j}$ is a non-degenerate random variable, $P(\xi_{j}=0)=0$. (2) $w_{j}=|\widetilde{\beta}_{j}^{ols}|^{-\gamma}=O_{p}(1)$.  
} \\

\noindent\emph{Proof of Lemma A2.} 
	Part (1) is the standard spurious regression result (Phillips, 1986), where the OLS estimator converges to a functional of independent Brownian motions rather than to 0. Part (2) follows from Part (1) and the continuous mapping theorem, as the limit $\xi_j$ has no mass at zero (Phillips, 1986).  \qed\\

\medskip
\noindent
\emph{Lemma A3 (score asymptotics)}. \emph{(1) Under the conditions of Lemma A1(1), for each $j \in S$, the normalized score $S_{n,j} \coloneqq \frac{1}{n}\sum_{t=1}^n x_{j,t}z_{t} \Rightarrow Z_{j}$, where $Z_{j}$ is non-degenerate continuous random variable,  $P(Z_{j}=0)=0$. (2) Under $\rho=1$ and (\textbf{X-NC}) assumed to hold, the score term $S_{n,j}=O_{p}(n)$ and $S_{n,j}/n$ has an $O_{p}(1)$ non-degenerate limit with no atom at 0 for each $j$}. \\
 
\noindent\emph{Proof of Lemma A3.}  Both parts follow straightforwardly from well-known results in the unit-root and cointegration literature (e.g., Phillips and Durlauf (1986)). \qed \\

\noindent
\emph{Proof of Proposition 1.} We proceed under the conditions collected in the statement of Proposition 1. \\

\noindent
\emph{Part a(i) - ($n$-consistency of the adaptive lasso estimator under (\textbf{X-IC}) and $|\rho|<1$):} 
Consider the adaptive lasso objective under the local parameterization ${\bm \beta}={\bm \beta}^{0}+n^{-1} {\bm u}$ for a generic vector ${\bm u}\in \mathbb{R}^{p}$. We let ${\bm u}=({\bm u}_{S}',{\bm u}_{S^{c}}')'$. 
Using $y_{t}={\bm x}_{t}'{\bm\beta}^{0}+z_{t}$ (recall that for notational simplicity our derivations ignore the intercept) and standard algebra, we write the objective function as
\begin{align}
	\mathcal{L}_n({\bm u}) &=  \sum_{t=1}^{n} (y_t - \bm x_t' (\bm \beta^0 + \frac{\bm u}{n}))^2 + \lambda_n \sum_{j=1}^{p} w_j |\beta_j^0 + \frac{u_j}{n}| \nonumber \\
	&= \sum_{t=1}^{n} (z_t - \frac{1}{n} \bm x_t' \bm u)^2 + \lambda_n \sum_{j=1}^{p} w_j |\beta_j^0 + \frac{u_j}{n}| \nonumber
\end{align}
and let $\hat{\bm u}^{AL} = n (\hat{\bm \beta}^{AL} - \bm \beta^0)$ be the minimizer of $\mathcal{L}_n({\bm u})$. 
Expanding the square in the first term gives
\begin{align}
	\mathcal{L}_n({\bm u}) &=  \sum_{t=1}^{n} \left( z_t^2 - 2 z_t \frac{1}{n} \bm x_t' \bm u + \frac{1}{n^2} (\bm x_t' \bm u)^2 \right) + \lambda_n \sum_{j=1}^{p} w_j \left|\beta_j^0 + \frac{u_j}{n}\right| \nonumber  \\
	&=\sum_{t=1}^{n} z_t^2 - \frac{2}{n} \sum_{t=1}^{n} z_t \bm x_t' \bm u + \frac{1}{n^2} \sum_{t=1}^{n} \bm u' \bm x_t \bm x_t' \bm u + \lambda_n \sum_{j=1}^{p} w_j \left|\beta_j^0 + \frac{u_j}{n}\right| \nonumber \\
	&= \sum_{t=1}^{n} z_t^2 - \frac{2}{n} \sum_{t=1}^{n} z_t \bm x_t' \bm u + \frac{1}{n^2} \bm u' \left( \sum_{t=1}^{n} \bm x_t \bm x_t' \right) \bm u + \lambda_n \sum_{j=1}^{p} w_j \left|\beta_j^0 + \frac{u_j}{n}\right| \nonumber
\end{align}
It is now also convenient to substract $\mathcal{L}_n(\bm 0)$ from $\mathcal{L}_n({\bm  u})$, consider
\begin{align}
	V_{n}({\bm u}) \coloneqq \mathcal{L}_n({\bm  u}) - \mathcal{L}_n({\bm 0}) &= \bm u' \left( \frac{\sum_{t=1}^{n} \bm x_t \bm x_t'}{n^{2}} \right) \bm u -  \frac{2}{n} \sum_{t=1}^{n} z_t \bm x_t' \bm u  + \lambda_n \sum_{j=1}^{p} w_j \left( \left|\beta_j^0 + \frac{u_j}{n}\right| - |\beta_j^0| \right) \nonumber \\
	& =  \underbrace{\bm u' \bm A_n \bm u - 2 \bm u' \bm B_n}_{Q_n(\bm u)} + \underbrace{\lambda_n \sum_{j=1}^p w_j \left( \left|\beta_j^0 + \frac{u_j}{n}\right| - |\beta_j^0| \right)}_{\text{Pen}_n(\bm u)} \label{eq:eq23} 
\end{align}
and focus on the optimisation of $V_{n}(\bm u)$. By standard FCLT results, $\bm A_n \Rightarrow \bm A$ and $\bm B_n \Rightarrow \bm B$.
The assumptions for part (a) of Proposition 1 state that $\bm x_{S,t}$ is not internally cointegrated and not cointegrated with $\bm x_{S^c,t}$. However, $\bm x_{S^c,t}$ may be internally cointegrated. While the full matrix $\bm A$ may therefore be singular due to internal cointegration within the inactive set $S^c$, the assumption of no cross-cointegration ensures that the stochastic trends in the active set $S$ are asymptotically linearly independent of those in $S^c$. This implies that the Schur complement of the active block is positive definite almost surely, i.e., $\bm A_{SS \cdot S^c} \coloneqq \bm A_{SS} - \bm A_{S S^c} (\bm A_{S^c S^c})^{-} \bm A_{S^c S} \succ 0$. Note that the absence of cross-cointegration between active and inactive predictors 
is key to ensure this property. Mathematically, it implies that the null space of ${\bm A}$ is entirely contained within the subspace spanned by the inactive parameters. This guarantees that the concentrated objective function remains strictly convex with respect to ${\bm u}_{S}$. 

\medskip 

\noindent We next formalize the asymptotic behaviour of ${V}_{n}({\bm u})$ by considering the two components in \eqref{eq:eq23} separately. 

\medskip
\noindent
\textit{Bounding the Quadratic Term:} we initially consider the behaviour of the quadratic term $Q_n(\bm u)$ with respect to $\bm u_S$. For any fixed $\bm u_{S^c}$, the quadratic form is minimized with respect to $\bm u_S$ when the gradient is zero. 
Due to the positive definiteness of $\bm A_{SS \cdot S^c}$, the profile quadratic function (concentrated on $\bm u_S$) grows quadratically:
\[
\min_{\bm u_{S^c}} \left( \bm u' \bm A \bm u \right) = \bm u_S' (\bm A_{SS \cdot S^c}) \bm u_S \ge c \|\bm u_S\|^2,
\]
for some $c > 0$. Intuitively, this ensures that regardless of the behavior of the inactive estimates, the active parameter estimates cannot diverge without pushing the objective function to infinity. \\

\noindent\textit{Bounding the Penalty Term:} We decompose the penalty component by active and inactive sets.
\begin{itemize}
	\item For $j \in S^c$ (Inactive): $\beta_j^0 = 0$. The term is $\lambda_n w_j |u_j|/n \ge 0$. We can drop this non-negative term for the lower bound.
	\item For $j \in S$ (Active): $\beta_j^0 \neq 0$. Using the reverse triangle inequality, the fact that $w_j \xrightarrow{p} \text{const}$ (Lemma A1), and invoking the assumption $\lambda_n/n \to 0$ it follows that 
	\[
	\text{Pen}_{n,S}(\bm u_S) \ge - \frac{\lambda_n}{n} \sum_{j \in S} w_j |u_j| = -o_p(1) \|\bm u_S\|_1.
	\]
\end{itemize}

\noindent Combining the bounds, for sufficiently large $n$ with high probability the objective function satisfies
\[
V_n(\bm u) \ge \bm u_S' \bm A_{SS \cdot S^c} \bm u_S - 2 \bm u' \bm B - o_p(1) \|\bm u_S\|.
\]
Since $\bm A_{SS \cdot S^c} \succ 0$ and $\bm B = O_p(1)$, the quadratic growth in $\bm u_S$ dominates both $-2\bm u' \bm B$ and the vanishing penalty term. As $\|\bm u_S\| \to \infty$, $V_n(\bm u) \to \infty$ regardless of $\bm u_{S^c}$. By convexity, the minimizer $\widehat{\bm u}_S^{AL} = n(\widehat{\bm\beta}^{AL}_S - \bm\beta^0_S)$ must therefore be stochastically bounded: $n(\widehat{\bm\beta}^{AL}_S - \bm\beta^0_S) = O_p(1).$ \\

\noindent
\emph{Part a(ii) - Sure-Screening Property:} The proof of the sure-screening property relies directly on the convergence rate established in Part (a)(i). It holds under \textbf{X-IC} (where inactives may be cointegrated) because, as established in Part (a)(i), the lack of cross-cointegration between actives and inactives ensures the active estimator $\widehat{\bm\beta}^{AL}_S$ remains consistent even if the inactive estimator is not unique. We aim to show that for every truly active variable $j \in S$, the probability of it being excluded from the model converges to zero, i.e., $P(\widehat{\beta}^{AL}_j = 0) \to 0$ as $n \to \infty$. From Part (a)(i), we established the super-consistency of the estimator on the active set which implies $\widehat{\beta}^{AL}_j \xrightarrow{p} \beta^0_j, \quad \forall j \in S.$ By the beta-min condition assumed in the proposition, the true coefficients are bounded away from zero: $\min_{j \in S} |\beta^0_j| \ge \underline{b} > 0.$ Consider the event where an active variable is dropped incorrectly, $\{ \widehat{\beta}^{AL}_j = 0 \}$. On this event, the estimation error must be at least as large as the signal
\[
\left| \widehat{\beta}^{AL}_j - \beta^0_j \right| = \left| 0 - \beta^0_j \right| = |\beta^0_j| \ge \underline{b}
\]
which implies $\{ \widehat{\beta}^{AL}_j = 0 \} \subseteq \{ |\widehat{\beta}^{AL}_j - \beta^0_j| \ge \underline{b} \}.$
Taking probabilities
\[
P(\widehat{\beta}^{AL}_j = 0) \le P\left( \left| \widehat{\beta}^{AL}_j - \beta^0_j \right| \ge \underline{b} \right).
\]

\noindent
Since $\widehat{\beta}^{AL}_j$ is consistent for $\beta^0_j$, the probability of the estimation error exceeding any fixed constant $\underline{b} > 0$ converges to zero. Thus, $P(\widehat{\beta}^{AL}_j = 0) \to 0$ for each $j \in S$.
Finally, since $|S|=s$ fixed, we apply the union bound to the event of missing \emph{any} active variable
\[
P(S \not\subseteq \widehat{S}) = P\left( \bigcup_{j \in S} \{ \widehat{\beta}^{AL}_j = 0 \} \right) \le \sum_{j \in S} P(\widehat{\beta}^{AL}_j = 0) \to 0.
\]
Consequently, $P(S \subseteq \widehat{S}) \to 1$. \\

\noindent
\emph{Part b(i) - $n$-consistency of the adaptive lasso estimator under (\textbf{X-NC}) and full model selection consistency:} We operate under (\textbf{X-NC}), where there is no cointegration among any components of $\bm x_t$. This implies that the limit matrix $\bm A$ is positive definite almost surely. 
We use the same objective function expansion $V_n(\bm u)$ as in Part (a)(i) but unlike in Part (a)(i) where we relied on a profile argument, here the quadratic form $\bm u' \bm A_n \bm u$ provides a strictly convex lower bound with respect to the \emph{entire} vector $\bm u$. We focus on bounding the penalty term to ensure that it does not destabilize the quadratic minimum. 
\begin{itemize}
	\item For active variables ($j \in S$), we use the reverse triangle inequality: $|\beta_j^0 + u_j/n| - |\beta_j^0| \ge -|u_j|/n$. Since $w_j = O_p(1)$ and $\lambda_n/n \to 0$, the penalty contribution is bounded below by $-(\lambda_n/n) w_j |u_j| = -o_p(1)|u_j|$.
	\item For inactive variables ($j \in S^c$), $\beta_j^0 = 0$, so the term is $\lambda_n w_j |u_j|/n \ge 0$. As the penalty term is nonnegative, it can be dropped for the purpose of establishing a lower bound.
\end{itemize}

\noindent
Combining these, for sufficiently large $n$ and with high probability
\[
V_n(\bm u) \ge \bm u' \bm A_n \bm u - 2 \bm u' \bm B_n - o_p(1)\|\bm u\|.
\]

\noindent
Let $\lambda_{\min}(\bm A)$ denote the minimum eigenvalue of the limit matrix. Since $\bm A \succ 0$, we have $\lambda_{\min}(\bm A) > c$ for some $c > 0$. Thus, the quadratic term grows as $c\|\bm u\|^2$, which asymptotically dominates the linear signal term ($2\bm u' \bm B_n$) and the vanishing penalty term.
Because the objective function diverges to $+\infty$ as $\|\bm u\| \to \infty$ and convex, the minimizer $\widehat{\bm u}$ must lie within a stochastically bounded set. This implies $\widehat{\bm u} = n(\widehat{\bm\beta}^{AL} - \bm\beta^0) = O_p(1)$.\\

\noindent
From Part (a)(ii) (Sure Screening), we know that $P(S \subseteq \widehat{S}) \to 1$. To prove full consistency, it suffices to show that no inactive variables are selected asymptotically, i.e., $P(j \in \widehat{S}) \to 0$ for all $j \in S^c$. Consider an inactive variable $j \in S^c$ (where $\beta^0_j = 0$). Suppose, for the sake of contradiction, that this variable is selected ($\widehat{\beta}^{AL}_j \neq 0$). The KKT stationarity condition must hold:
\begin{equation}
	\left| \mathcal{S}_{n,j}(\widehat{\bm\beta}^{AL}) \right| = \frac{\lambda_n}{n} w_j. \label{eq:eq24}
\end{equation}
\noindent
We analyze the asymptotic order of magnitude of both sides.\\

\noindent
\textit{Right-Hand Side (Penalty):} Under \textbf{X-NC}, Lemma A1(2) applies. The initial OLS estimator satisfies $n\widetilde{\beta}_j^{ols} = O_p(1)$ (non-degenerate limit), which implies that the weights explode at rate $n^\gamma$ and $w_j = n^\gamma O_p(1).$ Substituting this into the penalty term 
\[
\text{RHS} = \frac{\lambda_n}{n} O_p(n^\gamma) = O_p(\lambda_n n^{\gamma-1}).
\]
By the assumption $\lambda_n n^{\gamma-1} \to \infty$, the RHS diverges to infinity in probability.\\

\noindent
\textit{Left-Hand Side (Score):} Using the score decomposition:
\begin{equation}
	\mathcal{S}_{n,j}(\widehat{\bm\beta}^{AL}) = \underbrace{\frac{2}{n} \sum_{t=1}^n x_{tj} z_t}_{\text{Term 1}} - \underbrace{\frac{2}{n} \sum_{t=1}^n x_{tj} \bm x_t' (\widehat{\bm\beta}^{AL} - \bm\beta^0)}_{\text{Term 2}}.
	\label{eq:eq25}
\end{equation}
\begin{itemize}
	\item \textbf{Term 1:} By Lemma A2, the normalized score evaluated at the true parameter involves the cross-product of I(1) regressors and stationary I(0) errors. By the FCLT, it converges to a well-defined random variable
	\[
	\frac{2}{n} \sum_{t=1}^n x_{tj} z_t = O_p(1).
	\]
	\item \textbf{Term 2:} We use the convergence rate established in Step (i) above. We have $n(\widehat{\bm\beta}^{AL} - \bm\beta^0) = O_p(1)$. The Gram matrix of regressors satisfies $n^{-2} \sum \bm x_t \bm x_t' = O_p(1)$. Thus:
	\[
	\frac{2}{n} \sum_{t=1}^n x_{tj} \bm x_t' (\widehat{\bm\beta}^{AL} - \bm\beta^0) = 2 \left( \frac{1}{n^2} \sum_{t=1}^n x_{tj} \bm x_t' \right) n(\widehat{\bm\beta}^{AL} - \bm\beta^0) = O_p(1) \cdot O_p(1) = O_p(1).
	\]
\end{itemize}

\noindent
Combining these terms, the Left-Hand Side is stochastically bounded: $\text{LHS} = O_p(1).$ Thus the KKT equality \eqref{eq:eq24} implies an equality between a bounded term ($O_p(1)$) and a diverging term ($O_p(\lambda_n n^{\gamma-1}) \to \infty$).
\[
P\left( |\mathcal{S}_{n,j}| = \frac{\lambda_n}{n} w_j \right) \le P\left( O_p(1) \ge \text{diverging} \right) \to 0.
\]
Therefore, the probability that an inactive variable is selected approaches zero. Since $S^c$ is a fixed finite set, $P(\widehat{S} \cap S^c = \emptyset) \to 1$. Combined with the sure screening property ($P(S \subseteq \widehat{S}) \to 1$), we conclude: $P(\widehat{S} = S) \to 1.$ \qed \\

\noindent
\emph{Proof of Proposition 2.} We analyze the asymptotic behavior of the Adaptive lasso objective function scaled by $n^{-2}$. As we ignore the inclusion of the intercept, the estimator is defined as:
\begin{align}
	\widehat{\bm \beta}^{AL} & = \arg\min_{\bm \beta}  \sum_{t=1}^n (y_t - \bm x_t' \bm \beta)^2 + \lambda_n \sum_{j=1}^p w_j |\beta_j|. \label{eq:eq26}
\end{align}
Since $\rho=1$, $y_t$ is an $I(1)$ process. By assumption, $\bm x_t$ contains $p$ non-cointegrated $I(1)$ processes. We recall that the standard FCLT implies the following joint weak convergence as $n \to \infty$:
\begin{align}
	\frac{1}{n^2} \sum_{t=1}^n \bm x_t \bm x_t' &\Rightarrow \bm A \succ 0, 	\label{eq:eq27}\\
	\frac{1}{n^2} \sum_{t=1}^n \bm x_t y_t &\Rightarrow \bm b, 	\label{eq:eq28} \\
	\frac{1}{n^2} \sum_{t=1}^n y_t^2 &\Rightarrow c.  \label{eq:eq29}
\end{align}

		\medskip
		\noindent
		\emph{Part (i): Boundedness ($\widehat{\bm \beta}^{AL} = O_p(1)$).}
		Consider the rescaled objective function $\widetilde{Q}_n(\bm \beta) \coloneqq n^{-2} \sum_{t=1}^n (y_t - \bm x_t' \bm \beta)^2 + n^{-2} \lambda_n \sum_{j=1}^p w_j |\beta_j|$.
		Expanding the quadratic term
		\[
		\widetilde{Q}_n(\bm \beta) = \frac{1}{n^2}\sum_{t=1}^n y_t^2 - 2 \bm \beta' \left( \frac{1}{n^2}\sum_{t=1}^n \bm x_t y_t \right) + \bm \beta' \left( \frac{1}{n^2}\sum_{t=1}^n \bm x_t \bm x_t' \right) \bm \beta + \frac{\lambda_n}{n^2} \sum_{j=1}^p w_j |\beta_j|.
		\]
		We analyze the penalty term. From Lemma A2, under spurious regression, the OLS estimator $\widetilde{\beta}_j^{ols} = O_p(1)$ converges to a non-degenerate random variable. Thus, the weights $w_j = |\widetilde{\beta}_j^{ols}|^{-\gamma} = O_p(1)$.
		Given the assumption $\lambda_n / n \to 0$, it follows that $\lambda_n / n^2 \to 0$. Therefore
		for any fixed ${\bm \beta}$
		$$\text{Pen}_n(\bm \beta) = \left( \frac{\lambda_n}{n^2} \right) \sum_{j=1}^p O_p(1) |\beta_j| \xrightarrow{p} 0.$$
	Consequently, the objective function $\widetilde{Q}_{n}(\bm \beta)$ converges pointwise in distribution to the stochastic limit process $\widetilde{Q}(\bm \beta)$
			\[
		\widetilde{Q}_n(\bm \beta) \Rightarrow \widetilde{Q}(\bm \beta) \coloneqq c - 2 \bm \beta' \bm b + \bm \beta' \bm A \bm \beta.
		\]
		Since $\bm A \succ 0$ almost surely, the limit function $\widetilde{Q}(\bm \beta)$ is strictly convex with a unique minimizer $\bm \beta^* = \bm A^{-1} \bm b$.
		By convexity and the pointwise convergence to $\widetilde{Q}(\bm \beta)$, the Argmax Continuous Mapping Theorem implies that the minimizer of $\widetilde{Q}_n(\bm \beta)$ converges in distribution to the minimizer of $\widetilde{Q}$
		\[
		\widehat{\bm \beta}^{AL} \Rightarrow \bm \beta^*.
		\]
		Since $\bm \beta^*$ is a well-defined random variable (a functional of Brownian motions), it is stochastically bounded. Thus, $\widehat{\bm \beta}^{AL} = O_p(1)$.
		
		\medskip
		\noindent
		\emph{Part (ii): Selection Failure ($P(\widehat{\beta}^{AL}_j = 0) \to 0$).} 
		We establish that the adaptive lasso estimator asymptotically collapses to the unpenalized OLS estimator, which is non-zero with probability 1.
		The unpenalized OLS estimator is given by $\widetilde{{\bm \beta}}^{ols} = (\sum \bm x_t \bm x_t')^{-1} \sum \bm x_t y_t$. Using the limits in \eqref{eq:eq27} and \eqref{eq:eq28}, $\widetilde{\bm \beta}^{ols} \Rightarrow \bm A^{-1} \bm b$.
		The KKT optimality condition for the adaptive lasso estimator $\widehat{\bm \beta}^{AL}$ is:
		\[
		2 \sum_{t=1}^n \bm x_t (y_t - \bm x_t' \widehat{\bm \beta}^{AL}) = \lambda_n \bm W \bm s,
		\]
		where $\bm W = \text{diag}(w_1, \dots, w_p)$ and $\bm s$ is the subgradient vector with elements $s_j \in \text{sgn}(\widehat{\beta}_j^{AL})$.
		Rearranging terms
		\[
		2 \sum_{t=1}^n \bm x_t y_t - 2 \left( \sum_{t=1}^n \bm x_t \bm x_t' \right) \widehat{\bm \beta}^{AL} = \lambda_n \bm W \bm s.
		\]
	\noindent
	Given the positive definiteness of the limit matrix ${\bm A}$, for sufficiently large $n$ we can pre-multiply by the inverse of the Gram matrix  $(\sum \bm x_t \bm x_t')^{-1}$
		\[
		2 \widetilde{\bm \beta}^{ols} - 2 \widehat{\bm \beta}^{AL} = \lambda_n \left( \sum_{t=1}^n \bm x_t \bm x_t' \right)^{-1} \bm W \bm s.
		\]
		Solving for the difference between the estimators gives
		\begin{equation}
			\widehat{\bm \beta}^{AL} - \widetilde{\bm \beta}^{ols} = -\frac{1}{2} \left( \frac{1}{n^2} \sum_{t=1}^n \bm x_t \bm x_t' \right)^{-1} \frac{\lambda_n}{n^2} \bm W \bm s.  \label{eq:eq30}
		\end{equation}
		We analyze the order of the right-hand side of \eqref{eq:eq30}: $( n^{-2} \sum \bm x_t \bm x_t' )^{-1} \Rightarrow \bm A^{-1} = O_p(1)$, $\bm W = O_p(1)$ (as established in Lemma A3), $\bm s$ is bounded by 1 in element-wise magnitude. The scalar factor is $\lambda_n / n^2$. Since $\lambda_n / n \to 0$, it follows that $\lambda_n / n^2 \to 0$. \\
		
		\noindent
		Consequently, the difference vanishes asymptotically:
		\[
		\widehat{\bm \beta}^{AL} - \widetilde{\bm \beta}^{ols} = O_p(1) \cdot o(1) \cdot O_p(1) = o_p(1).
		\]
		
		\medskip
		\noindent
		Next, let $\xi_j$ denote the $j$-th element of the limit random vector $\bm A^{-1} \bm b$. The distribution of $\xi_j$ is a ratio of functionals of Brownian motions and is absolutely continuous. Therefore, $P(\xi_j = 0) = 0$.
		Since $\widehat{\beta}^{AL}_j \xrightarrow{p} \xi_j$ and $P(\xi_j = 0) = 0$, it follows that the probability of the estimator being exactly zero converges to zero: $P(\widehat{\beta}^{AL}_j = 0) \to 0$. Thus the sparse penalty is asymptotically insufficient to force any coefficient to zero against the spurious trend. Taking the intersection over the fixed number of predictors $p$, we conclude $P(\widehat{S} = \{1, \dots, p\}) \to 1$. \qed \\
		
	\noindent
	\noindent
	\emph{Remark A1.} It is instructive to contrast the failure in Proposition 2 with the success in Proposition 1. In the cointegrated setting (Proposition 1), model selection consistency is achieved provided $\lambda_n n^{\gamma-1} \to \infty$. This condition ensures the penalty term dominates the score term, which is $O_p(n)$ (driven by $I(1) \cdot I(0)$ correlations), because the weights explode at rate $n^\gamma$. In the spurious regression setting (Proposition 2), the context changes fundamentally in two ways. First, the score term becomes $O_p(n^2)$ (driven by the spurious correlation of independent $I(1)$ processes). Second,  the initial OLS estimator is inconsistent, converging to a random variable rather than zero. Consequently, the adaptive weights do not explode but remain $O_p(1)$. As a result, satisfying the condition $\lambda_n n^{\gamma-1} \to \infty$ is insufficient to suppress the noise. To force the inactive coefficients to zero in this regime, the penalty parameter would need to grow faster than the spurious trend itself (i.e., $\lambda_n > O_p(n^2)$), a rate that is incompatible with the standard tuning assumption $\lambda_n = o(n)$.  \\

We now proceed with the Proof of Proposition 3. A key feature of the proposed model selection based approach for detecting sparse cointegration is its robustness to the ``internal'' cointegration properties of the ${\bm x}_{t}'s$. We initially introduce an additional lemma (Lemma A4) designed to establish the fact that under both $|\rho|<1$ and $\rho=1$ the adaptive lasso residuals 
$\hat{z}_{t}$ share the same I(0)/I(1)'ness properties as the $z_{t}'s$ themselves. \\

\noindent
\emph{Lemma A4 (Universal Residual Behavior).}
\emph{Let $\hat{z}_t$ be the adaptive lasso residuals. Regardless of internal cointegration in $\bm x_t$:}
(i) \emph{If $|\rho|<1$ (cointegration): $\frac{1}{n} \sum \hat{z}_t^2 = O_p(1)$.} (ii) \emph{If $\rho=1$ (spurious): $\frac{1}{n^2} \sum \hat{z}_t^2 \Rightarrow V > 0$.} \\

\noindent
\emph{Proof of Lemma A4.} The adaptive lasso estimator $\widehat{\bm \beta}^{AL}$ is defined as the global minimizer of the penalized objective function. Consequently, the value of the objective function at the estimator must be less than or equal to its value at any other point, including the true parameter vector $\bm \beta^0$
$$\sum_{t=1}^n \hat{z}_t^2 + \lambda_n \sum_{j=1}^p w_j |\widehat{\beta}_j^{AL}| \leq \sum_{t=1}^n (y_t - \bm x_t' \bm \beta^0)^2 + \lambda_n \sum_{j=1}^p w_j |\beta_j^0|.$$
\noindent
Dropping the non-negative penalty term on the LHS, we obtain the upper bound for the RSS
\[
\sum_{t=1}^n \hat{z}_t^2 \leq \sum_{t=1}^n (y_t - \bm x_t' \bm \beta^0)^2 + \lambda_n \sum w_j |\beta_j^0|.
\]

\noindent 
We next analyze the order of the RHS terms: 

\medskip

\noindent
\textit{Part (i)}: Under cointegration ($|\rho|<1$), $y_t - \bm x_t' \bm \beta^0 = z_t$, which is a stationary $I(0)$ process. The sum of squares of a stationary process satisfies $\sum z_t^2 = O_p(n)$. For the penalty, recall that for active parameters ($j \in S$), the weights converge to constants $w_j = O_p(1)$. Since $\beta_j^0$ is fixed and $\lambda_n/n \to 0$, the penalty term is $O_p(\lambda_n) = o_p(n)$.
Thus, the total upper bound is $O_p(n) + o_p(n) = O_p(n)$. Dividing by $n$ gives
$$\frac{1}{n} \sum_{t=1}^n \hat{z}_t^2 \leq O_p(1).$$
This bound holds universally because even if $\bm \beta^0$ is not unique (due to rank deficiency), there exists at least one valid cointegrating vector for which this inequality holds.

\noindent
\textit{Part (ii)}: Under spurious regression ($\rho=1$), $y_t$ is an $I(1)$ process.
By the Functional Central Limit Theorem (FCLT), $n^{-1/2} y_{\lfloor nr \rfloor} \Rightarrow B_y(r)$ and also $n^{-1/2} \bm x_{\lfloor nr \rfloor} \Rightarrow \bm B_x(r)$, where $B_y$ is a scalar Brownian motion and $\bm B_x$ is a vector Brownian motion.
From Proposition 2, we know the estimator is stochastically bounded, $\widehat{\bm \beta}^{AL} = O_p(1)$, and converges in distribution to some random vector $\bm \xi$.
The residuals satisfy
\[
\frac{\hat{z}_{\lfloor nr \rfloor}}{\sqrt{n}} = \frac{y_{\lfloor nr \rfloor}}{\sqrt{n}} - \left(\frac{\bm x_{\lfloor nr \rfloor}}{\sqrt{n}}\right)' \widehat{\bm \beta}^{AL} \Rightarrow B_y(r) - \bm B_x(r)' \bm \xi \coloneqq Q(r)
\]
and the limit process $Q(r)$ is non-zero with probability 1.
By the Continuous Mapping Theorem, the normalized sum of squared residuals converges to the integral of this non-zero process
\[
\frac{1}{n^2} \sum_{t=1}^n \hat{z}_t^2 = \frac{1}{n} \sum_{t=1}^n \left( \frac{\hat{z}_t}{\sqrt{n}} \right)^2 \Rightarrow \int_0^1 Q(r)^2 dr > 0.
\]
Thus, the residual variance scales with $n^2$, confirming that $\hat{z}_t$ remains $I(1)$. \qed \\

\noindent
\emph{Proof of Proposition 3.}
Let $\overline{\sigma}^{2}_{0} = n^{-1} \sum_{t} (\Delta \hat{z}_t - \hat{\bm \beta}_{0}'\boldsymbol{w}_t)^2$ and $\overline{\sigma}^{2}_{1} = n^{-1} \sum_{t} (\Delta \hat{z}_t - \hat{\phi}\hat{z}_{t-1} - \hat{\bm \beta}_{1}'\boldsymbol{w}_t)^2$ denote the residual variance estimators under the restricted (${\cal M}_0$, $\phi=0$) and unrestricted (${\cal M}_1$, $\phi \neq 0$) specifications, respectively. The difference in information criteria is defined as
\begin{equation}
	\Delta IC_n(\hat{z}) = \ln\left(\overline{\sigma}^{2}_{0}\right) - \ln\left(\overline{\sigma}^{2}_{1}\right) - \frac{c_n}{n} = \ln\left( \frac{\overline{\sigma}^{2}_{0}}{\overline{\sigma}^{2}_{1}} \right) - \frac{c_n}{n}. \label{eq:eq31}
\end{equation}

\medskip
\noindent
\emph{Part (i) - Cointegration ($|\rho| < 1$).} 
By Lemma A4(i), the first-stage residuals satisfy $n^{-1}\sum \hat{z}_t^2 = O_p(1)$. They behave asymptotically as the stationary error process $z_t$.
Consider the auxiliary ADF regression. Under the alternative hypothesis of cointegration, the true coefficient on the lagged level is $\phi = \rho - 1 < 0$.
The unrestricted model ${\cal M}_1$ produces a consistent estimator $\hat{\phi} \xrightarrow{p} \phi \neq 0$. Consequently, the estimated error variance converges to the variance of the innovations: $\overline{\sigma}^{2}_{1} \xrightarrow{p} \sigma_{\epsilon}^2$.
The restricted model ${\cal M}_0$ imposes $\phi=0$, which is a misspecification. The estimator minimizes the variance of the differences $\Delta \hat{z}_t$ without utilizing the mean-reverting level component $\hat{z}_{t-1}$. Thus, $\overline{\sigma}^{2}_{0} \xrightarrow{p} \sigma_{\Delta z}^2 > \sigma_{\epsilon}^2$.
The ratio of the variance estimators converges to a constant strictly greater than 1
\[
\frac{\overline{\sigma}^{2}_{0}}{\overline{\sigma}^{2}_{1}} \xrightarrow{p} \frac{\sigma_{\Delta z}^2}{\sigma_{\epsilon}^2} \coloneqq \kappa > 1.
\]
Substituting into \eqref{eq:eq31}:
\[
\Delta IC_n(\hat{z}) \xrightarrow{p} \ln(\kappa) - 0 > 0.
\]
Thus, the probability of selecting the unrestricted model approaches 1.

\medskip
\noindent
\emph{Part (ii): Spurious Regression ($\rho = 1$).}
By Lemma A4(ii), the first-stage residuals are $I(1)$ such that $n^{-2}\sum \hat{z}_t^2$ converges to a positive random variable. The dependent variable in the ADF regressions is $\Delta \hat{z}_t$, which is $I(0)$.
Under the null hypothesis of a unit root, the true parameter is $\phi = 0$.
The unrestricted model ${\cal M}_1$ estimates this zero parameter. The t-statistic for $\phi=0$, denoted as $t_{\phi}$, corresponds to a residual-based unit root statistic. From standard least-squares algebra 
\[
n\overline{\sigma}^{2}_{0} - n\overline{\sigma}^{2}_{1} = \frac{n\overline{\sigma}^{2}_{1}}{n-k-1} t_{\phi}^2 \approx \overline{\sigma}^{2}_{1} t_{\phi}^2.
\]
From conventional unit-root statistics, we have that $\overline{\sigma}^{2}_{1}=O_{p}(1)$ (the variance of the dependent variable $\Delta \hat{z}$ is stable) and we also know that $t_{\phi}^{2}=O_{p}(1)$ (see e.g. Phillips and Ouliaris (1990)). Rearranging for the ratio of variances we can therefore write 
\[
\frac{\overline{\sigma}^{2}_{0}}{\overline{\sigma}^{2}_{1}} = 1 + \frac{\overline{\sigma}^{2}_{0} - \overline{\sigma}^{2}_{1}}{\overline{\sigma}^{2}_{1}} = 1 + \frac{1}{n} \left( \frac{ RSS_0 - RSS_1 }{ \overline{\sigma}^{2}_{1} } \right) = 1 + \frac{O_p(1)}{n}.
\]
Using the Taylor expansion $\ln(1+x) = x + O(x^2)$ for small $x$:
\[
\ln\left( \frac{\overline{\sigma}^{2}_{0}}{\overline{\sigma}^{2}_{1}} \right) = \frac{O_p(1)}{n} = O_p(n^{-1}).
\]
Now consider the penalty term in \eqref{eq:eq31}:
\[
\Delta IC_n(\hat{z}) = O_p(n^{-1}) - \frac{c_n}{n} = \frac{1}{n} \left( \underbrace{O_p(1)}_{\text{Likelihood Ratio}} - \underbrace{c_n}_{\text{Penalty}} \right).
\]
Since $c_n \to \infty$ while the likelihood ratio term is $O_p(1)$, the term inside the parentheses becomes negative with probability approaching 1. Although the factor $1/n$ drives the magnitude of $\Delta IC_n$ to zero, it is strictly positive and does not alter the sign. Consequently, $\Delta IC_n(\hat{z}) < 0$ asymptotically, leading to the correct selection of the restricted (unit root) model.\qed

\end{document}